\documentclass{ptephy_v1}

\preprintnumber{XXXX-XXXX} 

\usepackage{wrapfig}

\begin{document}

\title{Canonical Nambu mechanics: relevance to string/M-theory and approaches to quantization}


\author{Tamiaki Yoneya}
\affil{Institute of Physics, University of Tokyo, Komaba, Meguro-ku, Tokyo 153-8902, Japan \email{tamyoneya@gmail.com}}





\begin{abstract}%
We review some aspects of Nambu mechanics 
on the basis of the works previously published 
separately by the present author. Main focuses  
are on three themes, its various symmetry structures, 
their possible relevance to string/M theory,  
and a Hamilton-Jacobi like reformulation. 
We try to elucidate  
the basic ideas, most of which were 
rooted in more or less the same 
ground, and to explain motivations 
behind these works 
from a unified and vantage viewpoint. 
Various unsolved 
questions are mentioned. We also include some 
historical account on the genesis of the 
Nambu mechanics,  
and discuss (in the Appendix) 
some parallelism of various ideas behind the 
Nambu's paper with Dirac's old works 
which are related to the description 
of vortical flows in terms of gauge potentials.
\end{abstract}

\subjectindex{A0, A1, A6, B0, B2, B8}

\maketitle

\section{Introduction}

I would like to start this review\footnote{Written version of an invited talk in the workshop ``Space-time topology behind formation of micro-macro 
magneto-vortical structure by Nambu mechanics", 
Osaka City University, Sept. 28--Oct.1, 2020. 
}
 first by presenting 
a brief comment on the historical genesis of our subject. 
It seems to me that Nambu's paper ``Generalized Hamiltonian 
Dynamics" (GHD)\cite{nambu1973}, published almost 
five decades ago in 
1973, is a true `singularity', occupying a special position 
in the history of theoretical physics. 
His other seminal works such as the 
ones on a dynamical model of 
elementary particles based on an analogy with 
the BCS theory of superconductivity, the discovery of the 
string 
interpretation of the Veneziano amplitude, and  
many other notable works  
have all been generated under close interactions 
with the environment of the 
contemporary developments of 
physics at those periods. This is evidenced by the fact that 
in these cases more or less similar works by 
other authors have been appearing independently 
and almost simultaneously. 

The case of the GHD, in contrast, seems entirely different. 
As far as I know, he himself had never mentioned 
this paper in his later research papers, except 
for some expository accounts or reminiscences. 
However, we can clearly see from his Acknowledgement 
in the paper that 
the generalization of Hamiltonian dynamics 
attempted in this work
 had been a theme which he had been devoting 
himself for more than twenty years to take shape into, 
since his 
early years at the Osaka City University. 
He expressed 
his gratitude to K. Fusimi, a well-known 
expert on statistical mechanics, for encouragement. 
In fact, we can also 
see a similar acknowledgement to Fusimi 
in another \cite{nambuising} of his earlier 
papers that discussed the Onsagar's rigorous
 solution of the two-dimensional Ising model. 
Nambu's method of simplifying the Onsager solution 
using creation-annihilation operators in this work was 
quite close to the Fusimi's independent work
 \cite{fusimi} on honeycomb lattice. 
Apparently, Nambu's keen interest on statistical 
mechanics seems to motivate the singular GHD work. 
We can also envisage from his later lectures, 
belonging to last few talks \cite{nambutalk} in his whole life, that 
his special interest on fluid mechanics had also 
been closely connected with his ideas leading to the 
GHD work. 

However, apart from all his personal motivations, 
it seems difficult to imagine any interaction 
between the theme of the GHD paper and the 
contemporary developments of physics 
in the 1960s and the 70s. That is 
why I have described the GHD paper as 
a `singularity'. There does not seem to be 
any necessity for the appearance of the GHD paper in 
1973. It would be no surprise if we could find 
the same Nambu equations of motion in some old literature  
written even a century ago.  
But perhaps unexpectedly even to Nambu himself, 
the GHD paper has gradually turned out to be relevant 
to some aspects of string/M theory and also to other 
disciplines of physics, as the readers of this special 
section will recoginize from 
the wide range of the contributors with different 
backgrounds. 
The `singularity' 
became a small but very impressive star shining 
in the sky of theoretical physics. 

Now I outline the contents of this paper. 
In section 2, I will first discuss 
some salient features of the Nambu mechanics,  
focusing on its symmetry properties that are 
important to applications to string/M theory 
and to its possible schemes of quantization. 
We shall also 
mention some unsolved problems. 
Then in section 3, I will discuss the possible relevance 
of Nambu-type structures   
to string and membrane theories. In its main part (subsection 3.3), 
I will review our old attempt toward the discretized 
version of the Nambu bracket.  
In section 4, I will review my previous attempt 
toward a `covariantization' of the 
M-theory from the standpoint of the so-called 
discrete light-cone quantization (DLCQ). Formally, this 
attempt has a similar flavor with the regularization 
of (super) membranes discussed in the previous section. 
But the physical meaning will be quite remote. 
Only the main 
crux of the construction 
will be presented succinctly, since giving a full 
discussion would become too intricate and long for 
this review. Then in the final section, I return to 
the original intension of Nambu in the GHD 
which aimed at a new quantum mechanical formulation 
of dynamics. 
I will summarize the main points of a possible path, 
a generalized Hamilton-Jacobi-like reformulation of 
the Nambu mechanics, 
toward its ``wave-mechanical" quantization, on 
the basis of my previous work. 
A general discussion 
on the nature of quantization of the Nambu mechanics 
will also be given. 

The purpose of the whole discussions in this review 
is not to repeat 
the previous publications, but to 
elucidate various ideas related to the Nambu mechanics,  
which may be rather foreign 
to most of readers and have been scattered 
in different papers, from a unified 
standpoint of the symmetry structure 
of the Nambu mechanics. Hopefully, that would be 
useful for interested readers before going to the 
original papers directly. 
In Appendix, I will give a historical account 
on an interesting parallelism between 
Dirac's old attempts and Nambu's ideas, 
for the purpose of stimulating interests 
in these almost forgotten works.

\section{Symmetry structure of the Nambu mechanics}

The usual Hamilton equations 
of motion describe an incompressible 
flow in phase space,
\begin{equation}
	\frac{dX}{dt}=\{H,X\}\equiv D^i(H)\partial_iX
	\label{poisson}
\end{equation}
where $D^i(H)\partial_i$ is the operator corresponding 
to the vector flow governed by a 
Hamiltonian $H=H(\xi)$, 
\begin{align}
	D^i(H)\equiv \epsilon^{ij}\partial_jH, 
	\quad \partial_iD^i=0. \label{Liouville}
\end{align}
Here, for simplicity, the 
phase space is taken to be two dimensions 
$(\xi^1,\xi^2)=(q,p)$ and the flow is 
area-preserving. Throughout the 
present paper, we assume the summation convention 
for repeated indices of the components of the 
phase-space coordinates, unless stated otherwise explicitly. 
The Nambu equations of motion are simply a natural 
extension of this structure to a 3-dimensional 
phase space $(\xi^1, \xi^2, \xi^3)$, by 
introducing two Hamiltonians $H, G$ such that
\begin{align}
	&\frac{dX}{dt}=\{H, G, X\}\equiv 
	D^i(H,G)\partial_iX, \label{Nflow}\\
	&D^i(H,G)\equiv \epsilon^{ijk}\partial_jH\partial_kG, 
	\quad 
	\partial_iD^i=0. \label{LiouvilleN}
	\end{align}
Thus the area-preserving flow is now replaced by 
the three dimensional volume-preserving flow, the 
Poisson bracket being extended to the 3-dimensional 
Jacobian,
\begin{align}
	\{K,L,M\}=\frac{\partial(K,L,M)}{\partial(\xi^1, 
	\xi^2,x^3)}=\epsilon^{ijk}\partial_iK\partial_jL
	\partial_kM. 
\end{align}
In the following, we call this expression ``Nambu bracket". 
As alluded to in the previous section, Nambu's original 
motivation for this generalization was statistical 
mechanics, where Liuoville theorem \eqref{Liouville} 
plays a critical role under the assumption of ergodicity. 
Accordingly, he suggested a canonical ensemble 
characterized by a generalized Boltzmann 
distribution with a weight factor $e^{-\beta H-\gamma G}$ 
with two temperature parameters $1/\beta$ and $1/\gamma$, 
corresponding to the two 
conserved quantities $H$ and $G$. 

It is obvious that the Nambu bracket can be 
extended to arbitrary $n$-dimensional phase space as 
\begin{align}
	\frac{dX}{dt}=\{H_1,H_2,\ldots,H_{n-1},X\}
	=D^i(H_1,H_2,\ldots,H_{n-1})\partial_iX, 
	\quad \partial_iD^i=0,
	\label{njacobian}
\end{align}
using the $n$ dimensional Jacobian, with $n-1$ 
conserved quantities $(H_1,H_2,
\ldots,H_{n-1})$. In the present paper, 
we treat only the case $n=3$, unless 
stated otherwise. 

As an example of physical systems possessing the above 
structure, Nambu mentioned the Euler equation 
for a rigid rotator, 
\begin{align}
	\frac{d\xi^1}{dt}=\frac{(I_2-I_3)\xi^2\xi^3}{I_2I_3}, 
	\quad 
	\frac{d\xi^2}{dt}=\frac{(I_3-I_1)\xi^3\xi^1}{I_3I_1}, 
	\quad 
	\frac{d\xi^3}{dt}=\frac{(I_1-I_2)\xi^1\xi^2}{I_1I_2},
	\label{Eulertop1}
\end{align}
\begin{align}
	\frac{d\xi^i}{dt}=\{H,G,\xi^i\}, 
	\quad H=\frac{1}{2}(\xi_1^2+\xi_2^2+\xi_3^2), 
	\quad 
	G=\frac{\xi_1^2}{2I_1}+\frac{\xi_2^2}{2I_2}
	+\frac{\xi_3^2}{2I_3}.
	\label{Eulertop2}
\end{align}
According to him, this example is enough as a justification 
for exploring the proposed idea. Indeed, it is 
remarkable that 
the phase space coordinates are identified directly 
with the components ($\xi^i=L_i$) of the 
angular momentum with 
respect to the principal axes of a rigid body and 
$I_i$'s are the corresponding moments of inertia. 
Thus the components of the angular momentum 
themselves are now the canonical coordinates satisfying 
the canonical Nambu bracket relation
\begin{align}
	\{L_i,L_j,L_k\}=\{\xi^i,\xi^j,\xi^k\}=\epsilon^{ijk}, 
	\label{canonicalNbracket}
\end{align}
instead of the standard Poisson bracket relation and 
the equations of motion
\begin{align}
	\{L_i,L_j\}=\epsilon_{ijk}L_k, 
	\quad 
	\frac{dL_i}{dt}=\{G,L_i\}
	\label{angularPB}
\end{align}
with the single Hamiltonian $G$. 
Note that, in terms of the Nambu bracket formulation, 
the conservation of both $H$ and $G$ is manifest. 
They  play 
completely symmetrical roles. By contrast, 
in the standard Poisson bracket formulation, 
$H$ of course plays the role of a Casimir 
invariant with respect to the 
algebra of the components of the angular momentum, 
while $G$ is the kinetic energy of rotation. 
The relation between these two different canonical 
structures will be clarified in section 5. 

\subsection{The fundamental identity}
With these preparations, let us now turn to the 
symmetry properties of this system. First we examine 
whether the canonical Nambu bracket relation 
$\{\xi^i, \xi^j, \xi^k\}=\epsilon^{ijk}$ 
is preserved under the flow \eqref{Nflow}. 
It requires that 
\begin{align}
	0
	=\{\{H,G,\xi^i\},\xi^j, \xi^k\}
	+\{\xi^i,\{H,G,\xi^j\},\xi^k\}
	+\{\xi^i,\xi^j,\{H,G,\xi^k\}\}.
	\label{flowinv}
\end{align}
This is guaranteed by the so-called ``fundamental identity" (FI) identity \cite{takh}
\begin{align}
	&\{F_1,F_2,\{F_3,F_4,F_5\}\}=\{\{F_1,F_2,F_3\},F_4,F_5\}\nonumber \\  
	&+\{F_3,\{F_1,F_2,F_4\},F_5\}+\{F_3,F_4,\{F_1,F_2,F_5\}\}, 
	\label{FI}
\end{align}
that is valid {\it by definition} 
for arbitrary five functions, $F_1,
F_2,\ldots,F_5$ of the coordinates $\xi^i$'s. 
By 
choosing $F_1=H, F_2=G$ and $F_3=\xi^i, F_4=\xi^j, 
F_5=\xi^k$ in this identity, 
\eqref{flowinv} immediately follows. 
Apparently, Nambu himself was not aware of the 
identity \eqref{FI}. 
Clearly, the FI replaces the Jacobi 
identity of the Poisson bracket. 

In connection with the FI, it is important to 
notice that a seemingly natural extension 
of a single triple coordinates $(\xi^1, \xi^2,\xi^3)$ to 
the multiple sets of $3N$ coordinates $(\xi^1_a, \xi^2_a,\xi^3_a)$ 
with $a=1,2,\ldots, N$, by 
defining 
\begin{align}
	\{H,G,F\}^{(N)}
	\equiv 
	\sum_{a=1}^N\frac{\partial(H,G,F)}
	{\partial(\xi^1_a,\xi^2_a,\xi^3_a)}, 
	\label{3N}
\end{align}
violate the FI in general, except for 
completely decoupled systems where 
the functions $(H,G,F)$ are 
linear sums of the contributions depending
only on a single triplet.
This is a crucial 
difference of the Nambu dynamics from the 
ordinary Hamiltonian dynamics. In this sense, 
the universality of the former seems to be 
largely diminished in comparison with the latter. 
However, we may equally take a different standpoint 
that the 
Nambu mechanics could be useful for 
some special cases in constraining 
applicable systems by higher symmetries, subject to 
which we next turn.

\subsection{{\rm N}-gauge symmetry}
Let us examine whether there are any freedom in
the choice of the pair of two Hamiltonians $(H,G)$. 
By rewriting the equations of motion explicitly 
for the coordinate $\xi^i$ using two-dimensional 
Jacobian as 
\begin{align}
	\frac{d\xi^i}{dt}=\frac{1}{2}\epsilon^{ijk}
	\frac{\partial(H,G)}{\partial(\xi^j, \xi^k)},
	\label{Neqjacobian}
\end{align} 
we find that, for a given set $(H,G)$, any different set 
of the new Hamiltonians $(H',G')$ 
satisfying 
\begin{align}
	\frac{\partial(H',G')}{\partial(H,G)}=1\quad 
	(\mbox{or} \quad 
	\epsilon^{ijk}\partial_jH\partial_kG
	=\epsilon^{ijk}\partial_jH'\partial_kG'\,\,)
\end{align}
gives the same equations of motion. 
This condition can equivalently be reformulated as  
\begin{align}
	H\partial_iG-H'
	\partial_iG'=
	\partial_i\Lambda
	\label{Ngauge}
\end{align}
for an arbitrary ``generating" function $\Lambda$, 
which means that $\Lambda$ can be regarded as 
a function of $(G,G')$ and 
\begin{align}
	\frac{\partial \Lambda}{\partial G}=H, \quad 
	\frac{\partial \Lambda}{\partial G'}=-H'.
	\end{align} 
This is reminiscent of the ordinary 
canonical transformations in the standard 
Hamiltonian formalism, at least in a formal sense. 
As noted originally by Nambu, this symmetry can 
alternatively be rephrased as a kind of 
gauge transformation. From a general viewpoint 
of volume-preserving flow, the vector field appearing 
$D^i(H,G)$ in \eqref{LiouvilleN} 
 can always be expressed by defining 
the gauge field $A_i$ through
\begin{align}
	D^i=\frac{1}{2}\epsilon^{ijk}
	F_{jk}, \quad F_{jk}=\partial_jA_k-
	\partial_kA_j.
	\label{gaugefield}
\end{align}
The expression \eqref{LiouvilleN} corresponds to 
the particular form (so-called Clebsch representation 
that is  
familiar in fluid mechanics) 
\begin{align}
	A_i=H\partial_iG+\partial_i\psi,
\end{align}
where $\psi$ is an arbitrary undetermined function. 
The transformation $(H,G)
\rightarrow (H',G')$ defined by \eqref{Ngauge} is nothing but a 
gauge transformation, 
\begin{align}
	\psi \rightarrow \psi-\Lambda,
\end{align}
that keeps the above particular form. 
Throughout the present paper, we call the 
transformation $(H,G)
\rightarrow (H',G')$ characterized by \eqref{Ngauge} the ``N-gauge" transformation. 

Nambu considered a further generalization 
of the equations of motion to 
\begin{align}
	\frac{dX}{dt}=\sum_a\{H_a,G_a,X\}
	\label{multipleHG}
\end{align}
by introducing an arbitrary number of the pairs $(H_a,G_a)$ of Hamiltonians. 
Then the N-gauge transformation is generalized to 
\begin{align}
	\sum_a\Bigl(H_a\partial_iG_a-H_a'\partial_iG_a')=
	\partial_i\Lambda
\end{align}
Hence, the N-gauge transformation 
becomes akin further to the ordinary 
canonical transformation for many variables as
\begin{align}
	\frac{\partial \Lambda}{\partial G_a}=H_a, 
	\quad 
	\frac{\partial \Lambda}{\partial G'_a}=-H'_a.
\end{align}
Then, however, none of $H_a$ or $G_a$ is  
conserved for a general choice of them 
unless $\{H_a, G_a, H_b\}=0=
\{H_a, G_a, G_b\}$ for all different sets 
with $a\ne b$. 
It would diminish the possible merit in  
adopting the Nambu bracket notation at least in the 
sense of a dynamical system, while from the 
standpoint of utilizing the Nambu flow toward 
a generalization of symmetry to higher ones in a given 
system it may still be useful. This viewpoint will be 
useful later.

\subsection{Genuine canonical transformation}
From the general viewpoint of the canonical structure 
of the equations of motion, the genuine canonical 
transformation of the Nambu equations of motion 
should be defined to be the coordinate 
transformations preserving the canonical 
bracket relation \eqref{canonicalNbracket}. 
Consequently, the flows described by the 
equations of motion are a special case 
of such general canonical transformations.\footnote{Nambu 
himself considered such canonical transformations which are 
linear with respect to $\xi^i$'s, 
and came to notice the difficulties 
mentioned above for the case with $3N$ variables related 
to \eqref{3N}. 
} 
Then the infinitesimal form of the canonical transformations with 
arbitrary two functions $(F, G)$ 
is given by
\begin{align}
	\delta \xi^i=\{F,G, \xi^i\}\equiv
	D(F,G)\xi^i
\end{align}
where
\begin{align}
	D(F,G)\equiv D^i(F,G)\partial_i.
\end{align}
The FI guarantees 
that the canonical bracket relation is 
preserved. 
We can then define a finite transformation by 
\begin{align}
	X(s)&\equiv \exp\big(s D(F,G)\big)
	\nonumber\\
	&=\sum_{n=0}^{\infty}\frac{s^n}{n!}\{F,G,\{F,G,\{
	\cdots, \{F,G,\{F,G,X\overbrace{\}\}\cdots, \}\}\}}^n
\end{align}
that satisfies the Nambu flow equation
\begin{align}
	\frac{dX(s)}{ds}=\{F,G,X(s)\}.
\end{align}
The group property of these 
transformations was discussed in our previous work \cite{almy}.  
First, using the FI, we find identically for any $X$, 
\begin{align}	
[D(F_1,G_1),D(F_2,G_2)]X&=\{F_1,G_1,\{F_2,G_2,X\}\}	-\{F_2,G_2,\{F_1,G_1,X\}\}	\nonumber \\	&=	\{\{F_1,G_1,F_2\},G_2,X\}+\{F_2,\{F_1,G_1,G_2\},X\}
\nonumber \\
&=D(\{F_1,G_1,F_2\},G_2)X+D(F_2,\{F_1,G_1,G_2\})X. 
\end{align}
But the last line must be antisymmetric with respect to 
$1$ and $2$ by definition. This allows us to conclude\footnote{There is a typo in eq. (2.19) in ref.\cite{almy}.}
\begin{align}
	[D(F_1,G_1),D(F_2,G_2)]
	=&
	\frac{1}{2}\Bigl(D(\{F_1,G_1,F_2\},G_2)+D(F_2,\{F_1,G_1,G_2\})	\Bigr.\nonumber \\
	&\Bigl.-D(\{F_2,G_2,F_1\},G_1)-D(F_1,\{F_2,G_2,G_1\})
	\Bigr).
\end{align}
Note that the right-hand side does not match with the naive expectation
\begin{align}
	[D(F_1,G_1),D(F_2,G_2)]=D(F_3,G_3)
	\nonumber
\end{align}
for some appropriate $F_3(F_1,G_1;F_2,G_2)$ and $G_3(F_1,G_1;
F_2,G_2)$. The reason behind this phenomena seems to be 
the fact that the Clebsch form for the gauge 
field is not the most general form for volume-preserving flows. This is in contrast to the 
area-preserving flows where the composition law is 
simply expressed as 
\begin{align}
 [D(H),D(G)]=D(K), \quad K=\{H,G\},
\end{align}
as a consequence of the Jacobi identity. 

Instead, we have pointed out in \cite{almy} that 
the following skew-symmetric triple commutator relation are  valid:
\begin{align}
	D(A_{[1},A_2) D(A_{3]_N},B)&=
	2D(\{A_1,A_2,A_3\},B), \\
	D(B_{[1},B_2)D(A_{[1},A_2)D(A_{3]_N},B_{3]_N})
	&=4D(\{A_1,A_2,A_3\},\{B_1,B_2,B_3\}),
\end{align}
where 
\begin{align}
	[A,B,C]_{{\rm N}}\equiv 
	ABC-ACB+BCA-BAC +CAB-CBA.
	\label{Ntriple}
\end{align}
This triple commutator satisfying a complete 
skew symmetry 
was originally defined by Nambu as a candidate 
for quantum version of the Nambu bracket. 
However, \eqref{Ntriple} does not satisfy 
the FI. It is not clear how to interpret 
this structure. 
It might be possible that the 
group-like property for the finite 
canonical transformations could
 be understood in terms of a new 
composition law which would lead to 
the above triple commutation law, as a special subset 
of the group of all possible volume-preserving flows. 
To my knowledge, this has never been 
realized in any closed form. More about the problems of 
finite canonical transformations will be 
discussed in section 4 in connection with 
a Hamilton-Jacobi like approach to 
the Nambu mechanics. 

\section{String/membrane theories and the Nambu bracket}

From the viewpoint of string theory, 
possible physical applications of the Nambu bracket 
and its associated symmetry arise  
in the theories of relativistic membranes 
that are expected to be relevant to the putative 
M-theory, which has been supposed to be the 
description of the strong coupling region of the 
type-IIA string theory. 
Before going to membranes, let us first 
recall the case of relativistic strings, 
where the Poisson bracket naturally plays 
a similar role. 

\subsection{The case of string and some generalizations}
A general form \cite{y1} for the classical action integral of 
the world sheet, parametrized by $(\xi^1,\xi^2)$ of a single string whose space-time coordinates 
are $X^{\mu}(\xi)$, 
takes the following form,
\begin{align}
	S_k=\int d^2\xi \, e
	\Bigl\{\frac{1}{e^k}\Bigl[
	-\frac{1}{2}(\sigma^{\mu\nu}
	\sigma_{\mu\nu})^2
	\Bigr]^{k/2}+k-1
	\Bigr\},
\end{align}
where $\sigma^{\mu\nu}\equiv \epsilon^{ij}
	\partial_iX^{\mu}\partial_jX^{\nu}$ is the 
	induced surface elements 
	of the world sheet and 
	 $e=e(\xi)$ is an auxiliary variable (``Ein-bein") 
defined on the 
world sheet whose role is to make 
the action parametrization invariant. Note that we use the unit where 
the slope parameter $\alpha'$ is equal to $1/4\pi$. 
For any positive integer $k$,  
this gives the same equations of motion upon 
eliminating the auxiliary variable $e$. 
The familiar Nambu-Goto action corresponds to 
the simplest case $k=1$. The next simplest 
case $k=2$ was 
proposed first by Schild \cite{Schild}, though he has not 
considered the re-parametrization invariance. This amounts  
to setting the gauge $e=1$ from the outset. 
For a demonstration of quantum-mechanical equivalence 
of this case to the so-called Polyakov action 
that uses the world-sheet metric tensor $h_{ij}$ 
as auxiliary fields in stead of the $e$, 
\begin{align}
	S_{{\rm P}}=-\frac{1}{2}	\int d^2\xi\sqrt{-h}\,h^{ij}
	\partial_iX^{\mu}
	\partial_jX_{\mu}, 
\end{align}
we refer readers to \cite{y1}. 

Now the surface element $\sigma^{\mu\nu}$ 
can be regarded as the Poisson bracket of 
$X^{\mu}$ and $X^{\nu}$ if we treat the 
world-sheet as a two-dimensional phase space 
as used in \eqref{poisson} of the previous section:
\begin{align}
	\sigma^{\mu\nu}=\{X^{\mu},X^{\nu}\}, 
\end{align}
in terms of which the $n=2$ action takes the form
\begin{align}
	S_2=-\int d^2\xi \Bigl[\frac{1}{2e}\{X^{\mu},X^{\nu}\}
	\{X_{\mu},X_{\nu}\}-e\Bigr].
	\label{Schildaction}
\end{align}
The equations of motion and the constraint associated with 
the $e$ are, respectively, 
\begin{align}
	\{X_{\mu},e^{-1}\{X^{\mu},X^{\nu}\}\}=0, 
	\quad	\frac{1}{2}\{X^{\mu}, X^{\nu}\}\{X_{\mu},X_{\nu}\}=-e^2. 
	\label{constarint1}
\end{align}
It should be noted that the 
constraint condition is actually 
equivalent to the standard Virasoro condition,  
conforming to the world-sheet conformal symmetry, 
\begin{align}
	P^2+ \partial_{\sigma}X^2=0, \quad 
	P\cdot \partial_{\sigma}X=0,
	\label{constraints2}
\end{align}
when we define the canonical momentum by 
\begin{align}
	P^{\mu}=\frac{1}{2e}\Bigl(\partial_{\tau}X^{\mu}(\partial_{\sigma}X)^2
	-\partial_{\sigma}X^{\mu}(\partial_{\tau}X
	\cdot \partial_{\sigma}X)\Bigr), 
\end{align}
where the time-like and the space-like 
coordinates in the sense of world sheet 
are denoted by $\tau$ and $\sigma$, respectively. 
This guarantees that the canonical formalism 
is equivalent to the ordinary one as 
derived from the Nambu-Goto action or the Polyakov action.

With the gauge choice $e=1$, the re-parametrization 
symmetry is reduced to area-preserving 
re-parametrizations. For this case, 
Nambu \cite{nambu2} has suggested 
an extension of the string theory by replacing 
the Poisson bracket $\sigma^{\mu\nu}$ by a 
commutator of covariant derivatives $D_{\mu}
=-i(\partial_{\mu}-iA_{\mu})$ of the 
non-Abelian gauge theory, 
$
\{X_{\mu},X_{\nu}\}\rightarrow -i[D_{\mu}, D_{\nu}].
$
To quote his own words, ``the string may be a special 
realization of gauge fields in which some dynamical 
degrees of freedom are frozen while the other 
have become classical in a sense." This seems to 
prophesy of various modern matrix models 
that are obtained by dimensional reductions  
from the maximal super Yang-Mills 
theory in 10 dimensions, and also 
of the discretized light-like gauge action for a relativistic membrane which was investigated first by Hoppe \cite{Hoppe} 
in his thesis (1982). 

On the other hand, in ref.\cite{y1}, the constraint equation 
in the form \eqref{constarint1} was interpreted 
as the classical 
form for an uncertainty relation of space-time 
$\Delta X\Delta T \gtrsim 1$, which the present author 
has been advocating as the qualitative but intrinsic characterization  \cite{yo3} 
of the short-distance space-time structure of (critical) string theory. 
For extensive discussions of this subject, I would like 
to refer the reader to \cite{yo2} 
and other earlier references therein. 

From the viewpoint of the emergence of 
the higher-dimensional phase space as in \eqref{njacobian}, 
it is also useful to rewrite the action integral 
for $D$-dimensional extended objects as 
\begin{align}
	S^{(d)}\equiv 
	\int d^D\xi\Bigl[
	p_{\mu_1\mu_2\cdots \mu_D}
	\{X^{\mu_1},X^{\mu_2},
	\ldots, X^{\mu_D}\}+
	\frac{e}{2}(p_{\mu_1\mu_2\cdots \mu_D}p^{\mu_1\mu_2\cdots \mu_D}+2)
	\Bigr], 
	\label{momentuaction}
\end{align}
where $p_{\mu_1\mu_2\cdots \mu_D}
$ is a new auxiliary ``momentum" variable. 
The first term suggests the $D$-dimensional 
form 
$
	d\omega^{(D)}=p_{\mu_1\mu_2\cdots \mu_n}
	dx^{\mu_1}\wedge dx^{\mu_2}\wedge 
	\cdots \wedge dx^{\mu_D}. 
$
However, the second term of the action integral is nothing but the Hamiltonian 
in the sense of the ordinary Hamiltonian dynamics with a 
single Hamiltonian, resulting in the Hamiltonian constraint
\begin{align}
	\frac{1}{2}p_{\mu_1\mu_2\cdots \mu_D}p^{\mu_1\mu_2\cdots \mu_D}=-1, 
\end{align}
which requires that the $D$-form 
$p_{\mu_1\mu_2\cdots \mu_D}$ must be {\it time}-like. 
The latter condition is again consistent with 
the interpretation of the space-{\it time} uncertainty relation 
which essentially governs the dynamics of these systems. 
This aspect of the space-time 
uncertainty relation (with $D=2$) has been discussed in detail in \cite{yo2} as a possible realization of non-commutative 
space-time geometry. 
The flow equation \eqref{njacobian} with $n-1$ $(=D-1)$
Hamiltonians 
should here be interpreted as describing symmetries that characterize
 this system, rather than the dynamical flow directly 
 which should correspond to the genuine time variable 
 $\tau$.  
 
 A lesson we should 
 learn from this form of the action is that 
 the appearance of the higher-dimensional phase-space 
 structure 
 does not necessarily imply the relevance of 
 the Nambu mechanics. Conversely, the Nambu-type 
 mechanics equipped with higher-dimensional phase 
 space does not directly imply the higher-dimensional 
 objects. For instance, the action 
 proposed by Takhtajan in his remarkable paper \cite{takh} is 
 formulated as if a one-dimensional 
string-like object with  ($\xi^i=\xi^i(\sigma,\tau)$) is treated. 
The presence of two Hamiltonians 
of Nambu mechanics corresponds to the 
Clebsch form $H\partial_{\sigma} Gd\sigma \wedge d\tau$,  
instead the second term in the above action \eqref{momentuaction}
with $D=2$ that involves no derivative. 
Because of this difference, the $\sigma$
direction in such a formulation 
merely parametrizes a continuous 
family of ordinary one-dimensional trajectories 
in three dimensional phase space $(\xi^1,\xi^2,\xi^3)$. 
Although the system has a re-prametrization 
symmetry with respect to $\sigma$ in a kinematical 
sense due to the $\sigma$-derivative, the $\sigma$ 
direction has no dynamical significance, 
since there is 
no Hamiltonian constraint for combining the 
$\sigma$ direction with the 
$\tau$ direction: the Hamiltonian constraint, 
if existed, 
 would be responsible 
for assigning a dynamical role to $\sigma$ of defining 
a potential energy. 

This crucial feature will become relevant and be elucidated 
 further from a slightly different perspective 
 to motivate an attempt 
toward a generalized Hamilton-Jacobi formalism 
of Nambu mechanics in section 5.

\subsection{Nambu bracket for the 
covariant regularization of a 
relativistic membrane}

Now let us focus our attention on the ($k=2, n=3$) case of the generalized Schild-type action \eqref{Schildaction},
\begin{align}
	S_{{\rm mem}}=-\int d^3\xi
	\Bigl(\frac{1}{2e}\{X^{\mu},X^{\nu},X^{\sigma}\}
	\{X_{\mu},X_{\nu},X_{\sigma}\}-e	\Bigr).
	\label{membranecov}
\end{align}
With the gauge choice $e=1$, the re-parametization 
symmetry reduce to the volume-preserving 
re-parametriazations. Unlike the case ($k=2,n=2$) 
of the string, 
the covariant quantization of this system turned out to be 
extremely difficult. Only tractable way even to this day is to 
adopt the light-like gauge condition $X^{\tau}=
X^{10}+X^0=\tau$. But still the system is 
non-linear: the Hamiltonian is 
\begin{align}
	H=\int d^2\sigma \frac{1}{P^+}
	\Bigl(p_i^2+\frac{1}{2}\{X^i,X^j\}^2\Bigr) 
	+\cdots
	\label{membhamilton}
\end{align}
where the indices $i,j,\ldots$ are 
SO(9) transverse directions and 
the $p_i$'s are the momentum variable conjugate 
to $X_i$'s. 
In the above, we have already alluded to the matrix regularization that was applied to this system. 
In \cite{dhn}, its supersymmetric 
generalization was studied in detail. 
The so-called BFSS model can be regarded as 
a re-interpretation of this system toward 
a possible non-perturbative formulation of M-theory 
in a special light-like frame. 

One of the hopes in the late 90s, perhaps pursued by 
many researchers, was to find appropriate 
discretized regularizations directly to \eqref{membranecov} 
without assuming the special light-like frame, 
such that it gives a tractable covariant formulation 
of the dynamics of (super) membranes. 
If we denote such a ``quantized version" of the 
Nambu bracket by $[X^{\mu}, X^{\nu}, X^{\sigma}]$ 
that could replace the Nambu bracket, the 
action would essentially be (in the $e=1$ gauge) 
\begin{align}
	S=\frac{1}{2}{\rm Tr}\Bigl([X^{\mu}, X^{\nu}, X^{\sigma}][X_{\mu}, X_{\nu}, X_{\sigma}]\Bigr) +\cdots, 
	\label{membraneactmat}
\end{align}
where $X^{\mu}, \ldots$ are now 
the discretized counterpart of the 
world-volume coordinates for which the symbol 
${\rm Tr}$ replaces the 
integral over the world volume.  
However, no one has achieved satisfactory 
progress in such attempts. 
For example, our work \cite{almy} was a by-product of an 
unsuccessful attempt along this line.

\subsection{Discretized versions 
of the Nambu bracket}

Here we briefly review some of the main results of \cite{almy}. 
If we restrict to the usual square matrices $A, B, C, 
\ldots$, we define
\begin{align}
	[A,B,C]\equiv ({\rm Tr}A)[B,C]+({\rm Tr}B)
	[C,A]+({\rm Tr}C)[A,B]. 
	\label{qNambumat}
\end{align}
Note that, 
when one of the matrices is the unit matrix, 
 this reduces to the usual commutator for 
 the remaining two matrices, and is obviously 
 totally skew symmetric and satisfies 
${\rm Tr}[A,B,C]=0$ for arbitrary three matrices.

Also it is not difficult to prove that the FI, 
\begin{align}
	[F,G,[A,B,C]]=[[F,G,A],B,C]+[A,[F,G,B],C]+[A,B,[F,G,C]],
\end{align}
is satisfied. 
First we calculate the l.h.s,
\begin{align}
	[F,G,[A,B,C]]&=({\rm Tr}A)[F,G,[B,C]]
	+({\rm c.p.})\nonumber \\
	&=({\rm Tr}A)({\rm Tr}F)[G,[B,C]]
	-({\rm Tr}A)({\rm Tr} G)[F,[B,C]]+({\rm c.p.})
	\label{FIlhs}
\end{align}
where the symbol  (c.p.) abbreviates the contributions 
obtained by the cyclic permutation of $A,B,C$ from the 
expressions appearing prior to it. 
The r.h.s of the FI is 
\begin{align}
	({\rm Tr}F)[[G,A],B,C]+({\rm Tr}G)[[A,F],B,C]
	+({\rm Tr}A)[[F,G],B,C]
	+{\rm (c.p.)}. 
\end{align}
The third term cancels after the corresponding 
contribution from (c.p.) are added. It is a 
consequence of 
the following identity,
\begin{align}
	({\rm Tr}A)[[F,G],B,C]+({\rm Tr}B)[[F,G],C,A]
	+({\rm Tr}C)[[F,G],A,B]=0. 
\end{align}
Thus the r.h.s of the FI reduces to 
\begin{align}
&({\rm Tr}F)[[G,A],B,C]+({\rm Tr}G)[[A,F],B,C]+({\rm c.p.})
\nonumber \\
&=({\rm Tr}F)\Bigl(({\rm Tr}B)[C,[G,A]]+({\rm Tr}C)[[G,A],B]
\Bigr)+({\rm Tr}G)\Bigl(({\rm Tr}B)[C,[A,F]]
+({\rm Tr}C)[[A,F],B]
\Bigr)\nonumber \\
&+({\rm c.p}). 
\label{FIrhs}
\end{align}
Now by comparing this result with \eqref{FIlhs}, it is equal 
to the latter owing to the Jacobi identity. For example, 
 the term $({\rm Tr}A)({\rm Tr}F)
[G,[B,C]]$ in \eqref{FIlhs} have 
the corresponding contribution 
$
({\rm Tr}F)({\rm Tr}A)([B,[G,C]]-[C, [G,B]]) 
$
in \eqref{FIrhs}, which indeed reduces to the former due to 
the Jacobi identity. 

Now let us examine the structure of the 
flow given by this realization of the discretized 
3-bracket with two bosonic matrices $(F,G)$:
\begin{align}
	\delta A\equiv i[F,G,A]=i[({\rm Tr}F)G-({\rm Tr}G)F,A]
	+i({\rm Tr}A)[F,G]
	\label{genegauge}
\end{align}
Note that the second term takes a very peculiar form,  
while the first one is nothing but the standard 
form of a gauge transformation. To the extent 
of being a matrix commutator, 
the second term allows us to make an unusual 
shift of the matrix independently of the 
original matrix $A$, provided that its trace is non-vanishing. 
Since $[F,G]$ with suitable choice of $(F,G)$ can be 
any element of the Cartan algebra of su($N$),
we can shift any Hermitian matrices with non-zero 
trace to the unit matrix: $A\rightarrow N^{-1}{\rm Tr}A$. 

On the other hand, the derivation law is {\it not} satisfied 
for generic products of matrices in the sense that 
\begin{align}
	[F,G,A]B+A[F,G,B]\ne [F,G,AB], 
\end{align}
owing to the presence of the second term in \eqref{genegauge}. 
However, provided that ${\rm Tr}A={\rm Tr}B=0$, 
\begin{align}
	\delta {\rm Tr}(AB)&\equiv 
	{\rm Tr}\Bigl((\delta A)B+A(\delta B)\Bigr) 
	\\
	&={\rm Tr}\Bigl(i[({\rm Tr}F)G-({\rm Tr}G)F,AB]\Bigr)=0,
	\label{inv} 
\end{align} 
which is valid for the trace of 
an arbitrary number of the products of traceless matrices. 
This result has a nontrivial significance for the 
symmetry of the action of the type \eqref{membraneactmat}, 
since the FI guarantees that 
\begin{align}
	\delta [X^{\mu},X^{\nu},X^{\sigma}]
	&\equiv [\delta X^{\mu},X^{\nu},X^{\nu}]
	+ [X^{\mu},\delta X^{\nu},X^{\sigma}]+ [X^{\mu},X^{\nu},\delta X^{\sigma}]\nonumber \\
	&=i[F,G,[X^{\mu},X^{\nu},X^{\sigma}]]=
	i[({\rm Tr}F)G-({\rm Tr}G)F, [X^{\mu},X^{\nu},X^{\sigma}]]
\end{align}
and ${\rm Tr}[X^{\mu},X^{\nu},X^{\sigma}]=0$. 
Thus the action \eqref{membraneactmat} is 
invariant under the generalized gauge transformation 
\eqref{genegauge}. 
These results show that the requirement of the 
derivation law in its most 
general form is not quite important 
from the viewpoint of symmetries, if one assumes that 
the physical observables are restricted 
exclusively to the objects 
that can be constructed as the polynomials of 
the `bracketed' objects. We can still have 
a useful set of invariants when the basic 
entities are subjected to the generalized gauge 
transformations. 

In ref.\cite{almy}, the possible extensions of these structures 
to the objects with 
three or many indices are also investigated. 
Here we only briefly mention the case 
of the cubic matrices $A_{pqm}, \ldots$ with three indices. 
We define the generalization of trace operations as
\begin{align}
	\langle A\rangle\equiv \sum_{pm}A_{pmp}, 
	\quad 
	\langle AB\rangle \equiv \sum_{pqm}A_{pmq}B_{qmp}, 
	\quad 
	\langle ABC\rangle\equiv \sum_{pqrm}
	A_{pmq}B_{qmr}C_{rmp}
\end{align}
that satisfy
\begin{align}
	\langle AB\rangle=\langle BA\rangle, 
	\quad 
	\langle ABC\rangle=\langle BCA \rangle=\langle CAB
	\rangle. 
\end{align}
A triple product is then defined as 
\begin{align}
	(ABC)_{ijk}=\sum_p A_{ijp}\langle B\rangle C_{pjk}
	=\sum_{pqm}A_{ijp}B_{qmq}C_{pjk}. 
\end{align}
In terms of them, the Nambu triple-bracket is 
\begin{align}
	[A,B,C]^{(3)}\equiv (ABC)+(BCA)+(CAB)-(CBA)-(ACB)-(BAC) 
\end{align}
that is skew-symmetric and satisfies the FI. 
Note that the middle index $j$ of $A_{ijk}$ in these 
definitions plays the role of internal index, while 
the two outer indices $i$ and $k$ behave like the indices of 
square matrices. Thus it is indeed natural to 
expect that the FI is satisfied by essentially the same 
mechanism as in the case of the square matrices with 
the definition \eqref{qNambumat}. 
For details of proof, we refer the reader to \cite{almy}, 
where the possibilities other than the above 
cubic products leading to the FI were mentioned. 
It is also straightforward to extend the cubic case 
to more indices. 

As in the case of the representation in terms of square matrices, the above triple brackets reduce to the commutator 
if at least one the three cubic matrices is 
a generalized {\it unit} cubic matrices defined by 
\begin{align}
	I_{ijk}\equiv \delta^{(j)}_{ik}=
	\begin{cases}
		0 &\,\,{\rm if}\,\, i\ne k \\
		1 &\,\,{\rm if}\,\, i=k
	\end{cases}
\end{align}
for any $j$. Then we have 
\begin{align}
	(AIB)_{ijk}=\langle I\rangle \sum_pA_{ijp}B_{pjk}, 
 \quad (IAB)=(BAI)=\langle A\rangle B, 
\end{align}
which leads to  
\begin{align}
	[A,I,B]^{(3)}_{ijk}=\langle I \rangle \sum_p(A_{ijp}B_{pjk}-B_{ijp}A_{pjk}).
\end{align}
Note that this can indeed be regarded as a commutator 
for any fixed middle index $j$. 

Furthermore, 
the trace for cubic case satisfies 
\begin{align}
	\langle [F,G,A]B\rangle+\langle A[F,G,B]\rangle=0
\end{align}
provided that $\langle A\rangle=\langle B\rangle=0$, 
which implies that the trace of the product of 
Nambu brackets $\langle [A,B,C][X,Y,Z]\rangle$ is 
invariant under the gauge transformation 
generated by the Nambu bracket, since $\langle A,B,C\rangle=0$ for arbitrary choices of 
cubic matrices $A,B,C$. The same is true for 
the generalized trace for the product of $n$ cubic matrices of the form
\begin{align}
	\langle AB\cdots Z\rangle
	\equiv \sum_{p_1,p_2,\cdots,p_n, m}
	A_{p_1mp_2}B_{p_2mp_3}
	\cdots Z_{p_nmp_1}.
\end{align}
Hence the trace of the products of arbitrary number 
of Nambu brackets are also gauge invariants. 

\section{An attempt toward a covariantized M(atrix) theory}

The properties reviewed in the previous section might 
be useful if one pursues the construction of covariant theories for membranes in a regularized form. 
In particular, in the case of the cubic matrices, 
the middle index is expected to play the role of the 
discretized time variable of the world volume, while the 
first and third indices would correspond to the 
two-dimensional discretized spatial coordinates. 
However, there is no known concrete attempt along this line. 
One among various apparent difficulties in achieving 
such a goal is 
the problem of extracting causal {\it dynamical} development 
with respect to this {\it would-be} discrete time. 
For example, a crucial problem is how to 
connect it to the classical action \eqref{membranecov}, 
where all the three parameters $\xi^i$ of the world-volume 
coordinates appear completely 
symmetrical manner, in the continuum limit.  
It seems rather difficult to imagine how 
such a symmetrical structure could naturally emerges from the 
asymmetrical roles of the three indices. From this 
viewpoint, it might be worthwhile to investigate the 
other possible definitions of triple products 
of cubic matrices suggested in \cite{almy}. 

In the present section, however, I will review a more modest 
approach which is trying to `covariantize' the 
 M(atrix) theory of BFSS \cite{bfss}, rather than insisting 
  further on 
the possibilities of direct regularized theory of 
membranes.  
This attempt \cite{yonecovmat} 
is based on a slight generalization 
of the definition of the 3-bracket in terms of the usual 
square matrices, for the purpose of implementing a 
new type of gauge 
symmetry, as has been already exhibited above in 
\eqref{genegauge} that is higher than the usual SU($N$)-type 
gauge symmetries. Since all the details have been 
published five years ago in the above reference, 
we will restrict ourselves only to main conceptual 
issues. I hope that the following concise 
summary would be useful for raising 
readers' attention to this old problem 
which seems to be almost forgotten for two decades up to now, 
in spite of its potential importance. 
I also wish to refer readers to \cite{yonecovmat} 
for a fuller bibliography related to this subject. 
 
\subsection{DLCQ approach to the M(atrix) theory 
and preparations toward fully convariant generalization}

Let us first briefly summarize what the DLCQ approach 
\cite{Suss} is. It has been proposed as a variant 
of the original M(atrix)-theory conjecture. According to 
the latter, the type IIA string theory in 10 space-time 
dimensions is interpreted as being compactified 
along a 
circle of radius $R_{11}=g_s\ell_s$ of 
the 10-th spatial direction ($x^{10}
\sim x^{10}+2\pi R_{11}$), where $g_s$ is the 
string coupling and 
$\ell_s$ $(\sim \sqrt{\alpha'})$ is the 
fundamental string length constant. 
The gravitational length in 11 dimensions is $\ell_{11}=g_s^{1/3}\ell_s$. 
The unit of 10-dimensional momentum is then equal to $
1/R_{11}$ and is interpreted as the mass of a single 
D0-brane (or `D-particle'). 
Now let us suppose that the total 
momentum $P^{\mu}$ of this system satisfies 
the mass-shell condition $P^{\mu}P_{\mu}+M_{{\rm eff}}^2=0$,  
where $M_{{\rm eff}}$ is the total effective mass, and 
consider the limit of 
large total 
10-th momentum $P_{10}=N/R_{11}$ $(
N\rightarrow \infty)$. This is the so-called infinite 
momentum frame (IMF) where we have the non-relativistic 
approximation, 
\begin{align}
	P^0=P^{10}+
	\sqrt{(P^i)^2+(P^{10})^2+M_{{\rm eff}}^2}-P^{10}
	\sim \frac{(P^i)^2+M_{{\rm eff}}^2}{2P^{10}}+
	\ldots
\end{align}
where the indices $i=1,2,\ldots, 9$ run only 
over the S0(9) 
directions. However, If we define the light-like 
momenta by $P^{\pm}=P^{10}\pm P^0$, the mass-shell 
condition is expressed exactly as 
\begin{align}
	-P^-=\frac{(P^i)^2+M_{{\rm eff}}^2}{P^+}.
\end{align}
for arbitrary value of $P^+$. 
In the IFM, we can make identification $P^+\sim 2P^{10}
\sim 2N/R_{11}$. 
The BFSS M(atrix) model was originally proposed 
on the basis of analogy of this phenomena with 
the membrane Hamiltonian \eqref{membhamilton} 
in the light-like gauge. The $M_{{\rm eff}}^2$ is 
replaced with the Hamiltonian 
for the effective (super) Yang-Mills theory 
for D0-branes, 
\begin{align}
	H=N{\rm Tr}\Bigl(\hat{\boldsymbol{P}}_i^2
	-\frac{1}{2\ell_{11}^4}[\boldsymbol{X}_i, 
	\boldsymbol{X}_j]^2 
	+\cdots\Bigr)
	\label{matrixhamilton}
\end{align} 
where $\hat{\boldsymbol{P}}_i$ ($i\in$ S0(9)) 
are the traceless 
components of the matrix momenta that are 
canonical conjugates to the traceless parts to the 
Higgs (hermitian $N\times N$)
 matrix fields $\boldsymbol{X}_i$: the diagonal 
parts of $\boldsymbol{X}_i$ represent 
the coordinates of $N$ D-particles while 
the off-diagonal parts correspond to 
short open strings connecting them. 
Intuitively, D-particles are 
nothing but `partons' as the constituents for membranes 
and, hopefully, other objects in M-theory. 
As a one-dimensional gauge theory with only time 
direction, the gauge field degrees of freedom 
do not appear and is signified only by the existence of 
a gauge constraint which is preserved by  
the time development described by the above 
Hamiltonian. 

Now the DLCQ approach is essentially 
a proposal of reinterpreting it 
by assuming the compactification 
along the light-like direction $x^- (\equiv x^{10}-x^0)
\sim x^-+2\pi R$ directly, which implies 
\begin{align}
	P^+=2N/R
	\label{Pplus}
\end{align}
for an arbitrary finite $N$ and $R$, instead of the 
compactification along the 10th spatial direction. 
In the large $N$ limit, both are formally equivalent to each 
other. There is, however, a crucial difference 
for the interpretation of Lorenz transformations. 
In the former, the original 
BFSS conjecture, a Lorentz boost along the 
10th spatial direction induces a discrete change of the 
quantum number $N$ with fixed and Lorentz invariant 
$R_{11}$. By contrast, in the DLCQ interpretation, 
a Lorentz boost is a continuous change of the 
radius parameter $R$ itself ($R\rightarrow e^{\rho}R, 
P^+\rightarrow e^{-\rho}P^+$), while by definition $N$ is 
Lorentz invariant and fixed. Thus, the longitudinal 
momentum $P^+$ is a genuine continuous dynamical 
variable. This also implies that the weak coupling 
limit $g_s\rightarrow 0$ is meaningful in the 
DLCQ interpretation and can be connected to the strong 
coupling limit by a large Lorentz boost. 
Of course, the limit of uncompactified 11-dimensional 
theory requires to take both $R$ and $N$ infinitely large. 

From these discussions, it seems clear that 
there must exist the Lorentz invariant formulation 
for $M_{{\rm eff}}^2$ 
as the generalization of \eqref{matrixhamilton} 
even for finite $N$, as a prerequisite for 
the feasibility of the DLCQ approach. 
Since now the theory must be meaningful as an exact 
formulation of, at least, one corner of
 M-theory even with finite and fixed $N$, the standpoint 
of interpreting the M(atrix) theory as a 
regularization of membrane should be abandoned 
if one seriously adopt the DLCQ point of view. 
This was the basic motivation 
for my previous work \cite{yonecovmat}. 
Namely, we should be able to extend the Hamiltonian 
to a Lorentz invariant $M_{{\rm eff}}^2$ in such a way 
that it reduces to 
\eqref{matrixhamilton} once we take the light-like frame. 
In order to realize this expectation, we have to 
require the followings:
\begin{enumerate}
\item Since the would-be Hamiltonian (as the 
mass-square operator) is itself Lorentz invariant, 
the time variable must also be Lorentz invariant 
as in the case of proper-time formalism for 
a relativistic particle quantum mechanics. 
\item  All of the 11-dimensional space-time directions 
must be treated on an equal footing as 
matrix degrees of freedom, since the transverse 
directions are already appearing as the matrix fields 
in the light-like gauge. 
\item It must 
be equipped with some higher-symmetries 
that encompass the ordinary SU($N$) gauge 
symmetry and enable us to eliminate the additional 
longitudinal matrix degrees of freedom by appropriate 
gauge fixing procedure. 
\end{enumerate}

Concerning the last point (3), the structure exhibited 
in \eqref{genegauge} seems promising, since it indeed 
encompasses the usual SU($N$) transformation by the 
presence of a non-standard shift term that can 
eliminate the traceless part of any single 
Hermitian matrix 
with non-vanishing trace. In fact, however, it is 
possible even to eliminate the restriction to a 
matrix with non-vanishing trace. 
We associate an auxiliary non-matrix variable  
$X_{{\rm M}}$ 
to a hermitian ($N\times N$) matrix $\boldsymbol{X}$, and 
denote the pair of 
them by $X=(X_{{\rm M}},\boldsymbol{X})$. 
Then we define the 3-bracket as 
\begin{align}
	[X,Y,Z]\equiv (0,X_{{\rm M}}[\boldsymbol{Y},
	\boldsymbol{Z}]+Y_{{\rm M}}[\boldsymbol{Z},
	\boldsymbol{X}]+Z_{{\rm M}}[\boldsymbol{X},
	\boldsymbol{Y}]).
\end{align}
Note that in this slight generalization of 
\eqref{qNambumat}, the role of the traces of the matrices 
are played by the `M'-components which are treated 
as new independent dynamical variables. 
It is easy to check that the FI is still satisfied 
in the same way as we have discussed in 
subsection 3.2, due essentially to 
the Jacobi identity for matrices. 
For example, the absence ({\it i.e.} $
[X,Y,Z]_{{\rm M}}=0$) of the 
M-component for the 3-bracket is guaranteed 
by the cancellation of the contributions 
involving the commutator $[\boldsymbol{F},
\boldsymbol{G}]$ {\it without} performing 
any trace operations for the three elements 
$(X,Y,Z)$ in the r.h.s of the FI,
\begin{align}
	[F,G,[X,Y,Z]]=[[F,G,X],Y,Z]+[X,[F,G,Y],Z]
	+[X,Y,[F,G,Z]].
	\label{FI2}
\end{align}
Otherwise we would have a contribution 
of the form $[X,Y,Z]_{{\rm M}}[\boldsymbol{F},
\boldsymbol{G}]$ which would ruin the 
consistency of the above definition. 

With this slight extension, the generalized 
gauge transformation takes the form
\begin{align}
	\delta X=i[F,G,X]=(0,i[F_{{\rm M}}
	\boldsymbol{G}-G_{{\rm M}}\boldsymbol{F}, 
	\boldsymbol{X}]+i[\boldsymbol{F},\boldsymbol{G}]
	X_{{\rm M}}).
	\label{3bragenetrans}
\end{align}
In this form, however, the SU($N$) part and the 
shift part are not yet completely independent 
to each other. This is remedied by introducing 
an arbitrary number of independent gauge 
variables $(F^a, G^a)$ discriminated by the 
indices $a=1,2,\ldots$, following 
Nambu's generalization \eqref{multipleHG}: 
\begin{align}
	\delta X=i\sum_a[F^a,G^a,X]=\sum_a(0,i[F^a_{{\rm M}}
	\boldsymbol{G}^a-G^a_{{\rm M}}\boldsymbol{F}^a, 
	\boldsymbol{X}]+i[\boldsymbol{F}^a,\boldsymbol{G}^a]
	X_{{\rm M}}),\label{multipleFG}
\end{align}
which can be redefined as 
\begin{align}
	&\delta_{HL}X\equiv \delta_{H}X+\delta_{L}X=(0,i[\boldsymbol{H},\boldsymbol{X}])+(0,\boldsymbol{L}X_{{\rm M}}),\\
	&\boldsymbol{H}\equiv \sum_a(F^a_{{\rm M}}
	\boldsymbol{G}^a-G^a_{{\rm M}}\boldsymbol{F}^a),
	\quad \boldsymbol{L}\equiv \sum_a i[\boldsymbol{F}^a,\boldsymbol{G}^a].
\end{align}
In this form, the matrix gauge variables 
$\boldsymbol{H}$ and $\boldsymbol{G}$, both of 
which are assumed to be traceless without 
losing generality, 
can be treated as two different 
Hermitian matrices that are completely independent 
of each other. Note that 
the distribution law for the 3-bracket under
 this gauge transformation, 
\begin{align}
	\delta_{HL}[X,Y,Z]=
	[\delta_{HL}X, Y,Z]+[X,\delta_{HL}Y, Z]+
	[X,Y,\delta_{HL}Z], 
\end{align}
is still satisfied after this 
extension, since each term with different $a$ of 
the r.h.s of \eqref{multipleFG} 
satisfies \eqref{FI2}, separately. 
In terms of this new notation, the original 
3-bracket form \eqref{3bragenetrans} is no more meaningful 
at least for the gauge 
transformation law, as has also been the case 
for the extended
 classical Nambu flow equations mentioned in the end 
of the subsection 2.2 with the multiple pairs $(H_a, G_a)$
instead of a single-pair of two-Hamiltonians.  
However, the 3-bracket notation will still 
be useful for describing the invariants
 (hence the action) under 
these generalized gauge transformations. 
For example, it is manifest, as an extension of \eqref{inv},  that 
\begin{align}
	\delta_{HL}\langle \prod_{i=1}^n[X_i,Y_i,Z_i]\rangle=0
\end{align}
where the symbol $\langle \prod_{i=1}^n A^{(i)}\rangle\equiv 
{\rm Tr}(\boldsymbol{A}^{(i)}\cdots)$ is used 
for the products of arbitrary number 
of the objects $A^{(i)}$ $(i=1,2,\ldots)$ whose M-components vanish, 
$A^{(i)}_{{\rm M}}=0$. 

\subsection{Covariantized M(atrix) theory}
Now I present my proposal for the covariantized 
M(atrix) theory satisfying all of the three requirements 
(1)$\sim$(3) above. To make the following account reasonably 
 short, 
only the final results for  
the bosonic part will be discussed. The further details,  including the 
fermionic part, the reader should refer to the 
original paper \cite{yonecovmat}. 
The basic degrees of freedom are the 11-dimensinal coordinate 
vectors $X^{\mu}=(X^{\mu}_{{\rm M}}, \boldsymbol{X}^{\mu})$ 
and the conjugate momentum vector 
$P^{\mu}=(P_{{\rm M}}^{\mu},\boldsymbol{P}^{\mu})$, 
both of which are functions of the single Lorentz-invariant 
proper time $\tau$. 
They are canonical conjugates to each other classically, 
in the sense of Lorentz-invariant canonical formalism 
with respect to $\tau$, satisfying 
the {\it usual} canonical (equal-time) Poisson bracket relations:
\begin{align}
	&\{X_{{\rm M}}^{\mu},P^{\nu}_{{\rm M}}\}_{{\rm P}}=
	\eta^{\mu\nu}, \\
	&\{X_{ab}^{\mu},P^{\nu}_{cd}\}_{{\rm P}}=
	\eta^{\mu\nu}\delta_{ad}\delta_{bc}, 
\end{align}
with all the other Poisson brackets being zero. In the second line, 
the indices $a,b, etc$ represent the matrix elements 
of the matrix part of the variables. 
The meaning and the role of the M-components $(X^{\mu}_{{\rm M}}, P^{\mu}_{{\rm M}})$ will be 
elucidated later. It should be kept in mind that 
all these variables are assumed to be scalar under the 
re-parametrization $\tau
\rightarrow \tau'$ with respect to the proper time $\tau$. 
In order to take into account the re-parametrization 
symmetry, we need to introduce 
the auxiliary ein-bein $e(\tau)$ satisfying 
$e(\tau)=(d\tau'/d\tau)e'(\tau)$. 

Throughout this subsection, when it becomes necessary to 
separate the center-of-mass degrees and the rest, we 
use the following convention: 
\begin{align}
	&X_{\circ}^{\mu}\equiv \frac{1}{N}{\rm Tr}(\boldsymbol{X}^{\mu}),\quad 
	\boldsymbol{X}^{\mu}=X_{\circ}^{\mu}+\hat{\boldsymbol{X}}^{\mu},\\
	&P_{\circ}^{\mu}\equiv {\rm Tr}(\boldsymbol{P}^{\mu}),
	\qquad 
	\boldsymbol{P}^{\mu}=\frac{1}{N}P_{\circ}^{\mu}
	+\hat{\boldsymbol{P}}^{\mu}. 
\end{align}
Thus any matrix with a hat is  traceless by definition. 
The gauge transformation law is 
\begin{align}
	\delta_{HL}\hat{\boldsymbol{P}}^{\mu}
	\equiv i[\boldsymbol{H},\hat{\boldsymbol{P}}^{\mu}]
	=\delta_H\hat{\boldsymbol{P}}^{\mu}, \quad
	\delta_{HL}P_{{\rm M}}^{\mu}=-{\rm Tr}(\boldsymbol{L}\boldsymbol{P}^{\mu})=
	\delta_LP_{{\rm M}}^{\mu}, 
\end{align}
with other components being inert. 
Together with the corresponding law 
for the coordinate variables 
$X^{\mu}=(X^{\mu}_{{\rm M}}, 
\boldsymbol{X}^{\mu})$ defined in the previous 
subsesction, 
the Poisson bracket relations 
are invariant. 

The total 
Poincar\'{e} integral 
corresponding to the canonical structure exhibited 
by the above Poisson bracket relations is 
given by
\begin{align}
	\int d\tau \Bigl[P_{{\rm M}\,\mu}
	\frac{dX_{{\rm M}}^{\mu}}{d\tau}+{\rm Tr}
	\Bigl(\boldsymbol{P}_{\mu}\frac{D\boldsymbol{X}^{\mu}}
	{D\tau}
	\Bigr)
	\Bigr]\equiv 
	\int d\tau \Bigl[P_{{\rm M}\,\mu}
	\frac{dX_{{\rm M}}^{\mu}}{d\tau}+
	P_{\circ\,\mu}\frac{DX_{\circ}^{\mu}}{D\tau}+{\rm Tr}
	\Bigl(\hat{\boldsymbol{P}}_{\mu}\frac{D\hat{\boldsymbol{X}}^{\mu}}
	{D\tau}
	\Bigr)
	\Bigr], 
\end{align}
where the derivative symbols with upper-case $D$ are the 
covariant derivatives, defined as follows. 
\begin{align}
&\frac{DX_{\circ}^{\mu}}{D\tau}= \frac{dX_{\circ}^{\mu}}{d\tau}-eB_{\circ}X_{{\rm M}}^{\mu}+e
{\rm Tr}(\boldsymbol{Z}\hat{\boldsymbol{X}}^{\mu}), \\
&\frac{D\hat{\boldsymbol{X}}^{\mu}}{D\tau}=
\frac{d\hat{\boldsymbol{X}}^{\mu}}{d\tau}+ie[\boldsymbol{A}, 
\boldsymbol{X}^{\mu}]-e\boldsymbol{B}X_{{\rm M}}^{\mu}, \\
&\frac{DP_{{\rm M}}^{\mu}}{D\tau}=\frac{dP_{{\rm M}}^{\mu}}{d\tau}
+e{\rm Tr}\bigl((\boldsymbol{B}+B_{\circ})\boldsymbol{P}^{\mu}\bigr)
=\frac{dP_{{\rm M}}^{\mu}}{d\tau}
+e{\rm Tr}(\boldsymbol{B}\boldsymbol{P}^{\mu})
+eB_{\circ}P_{\circ}^{\mu}, \\
&\frac{D\hat{\boldsymbol{P}}^{\mu}}{D\tau}=
\frac{d\hat{\boldsymbol{P}}^{\mu}}{d\tau}+ie
[\boldsymbol{A}, \boldsymbol{P}^{\mu}]
-e\boldsymbol{Z}P_{\circ}^{\mu}, 
\end{align}
It should be noted here that the $e$ is the ein-bein ({\it not}  
coupling constant), which is necessary to keep 
the reparametrization symmetry for the covariant 
derivatives. 
There are four gauge fields (the three traceless Hermitian 
matrices: $\boldsymbol{A}, 
\boldsymbol{B}, \boldsymbol{Z},$ and a scalar $B_{\circ}$) 
whose gauge transformations 
are, for $\delta_{HL}$,  
\begin{align}
	&\delta_{HL}\boldsymbol{A}=i[\boldsymbol{H},
	\boldsymbol{A}]-\frac{1}{e}\frac{d}{d\tau}\boldsymbol{H}\equiv -\frac{1}{e}
	\frac{D\boldsymbol{H}}{D\tau}, \\
	&\delta_{HL}\boldsymbol{B}=i[\boldsymbol{H},\boldsymbol{B}]
	-i[\boldsymbol{L},\boldsymbol{A}]+
	\frac{1}{e}\frac{D\boldsymbol{L}}{D\tau},\\
	&\delta_{HL}\boldsymbol{Z}=i[\boldsymbol{H},
	\boldsymbol{Z}],\\
	&\delta_{HL}B_{\circ}={\rm Tr}(\boldsymbol{L}
	\boldsymbol{Z}).
\end{align}
We have included the additional new 
gauge symmetry $\delta_w$ and $\delta_Y$, 
\begin{align}
	&\delta_wB_{\circ}=\frac{1}{e}\frac{dw}{d\tau}, 
	\quad \delta_w\boldsymbol{Z}=0, \\
	&\delta_YB _{\circ}=-{\rm Tr}(\boldsymbol{Y}
	\boldsymbol{B}), \quad 
	\delta_Y\boldsymbol{Z}=i[\boldsymbol{A},
	\boldsymbol{Z}]+\frac{1}{e}
	\frac{d\boldsymbol{Y}}{d\tau}\equiv \frac{1}{e}
	\frac{D\boldsymbol{Y}}{D\tau}, 
\end{align} 
for which the gauge transformation law of the canonical 
variables is given as
\begin{align}
	&\delta_wX_{\circ}^{\mu}=wX_{{\rm M}}^{\mu}, 
	\quad \delta_wP_{\circ}^{\mu}=0, \quad 
	\delta_wX_{{\rm M}}^{\mu}=0, 
	\quad \delta_wP^{\mu}_{{\rm M}}=-wP_{\circ}^{\mu}, \\
	&\delta_Y\hat{\boldsymbol{X}}^{\mu}=0, 
	\quad \delta_Y\hat{\boldsymbol{P}}^{\mu}=
	P_{\circ}^{\mu}\boldsymbol{Y}, \quad 
	\delta_YX_{\circ}^{\mu}=-{\rm Tr}(\boldsymbol{Y}
	\hat{\boldsymbol{X}}^{\mu}), 
	\quad \delta_YP^{\mu}_{\circ}=0, 
\end{align}
where $w$ and $\boldsymbol{Y}$ are an arbitrary function and an arbitrary traceless matrix functions, 
respectively, as parameters for the new gauge transformations. 
The canonical Poisson bracket relations are 
kept invariant under their actions. 
The other variables not shown explicitly are all 
inert. Note that 
the $P_{\circ}^{\mu}$ and $X_{{\rm M}}^{\mu}$ 
are themselves completely gauge invariant. 
The role of the covariant derivatives with these gauge 
fields is of course to make the transformations laws 
the same after acting these covariant derivative 
to the canonical variables. 
All of these gauge transformations are summarized by 
defining the canonical generator function as 
\begin{align}
	{\cal C}=wP_{\circ}\cdot X_{{\rm M}}
	+{\rm Tr}\Bigl(-(P_{\circ}\cdot \boldsymbol{X}
	)\boldsymbol{Y}+i\boldsymbol{P}_{\mu}
	[\boldsymbol{H},\boldsymbol{X}^{\mu}]
	+(X_{{\rm M}}\cdot\boldsymbol{P})\boldsymbol{L}\Bigr), 	
\end{align}
such that the transformation law are expressed by taking 
Poisson brackets between the ${\cal C}$ and 
the canonical variables. Here for brevity, 
some of the scalar products are represented by 
using dot symbols. In Fig.1, the structure 
of the gauge symmetries as a whole is summarized. 

\begin{figure}[htbp]
\begin{minipage}[t]{0.45\textwidth}
\begin{center}
\includegraphics[width=0.95\textwidth,clip]
{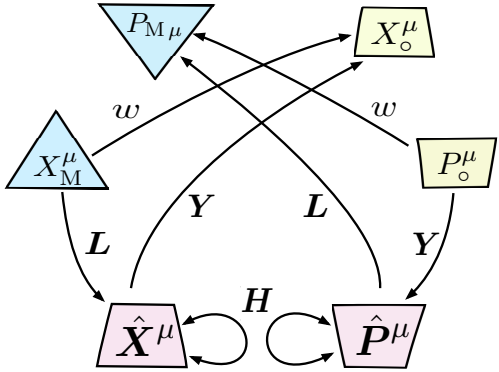}
\end{center}
\end{minipage}
\hspace{0.5cm}
\begin{minipage}[t]{0.52\textwidth}
\vspace{-5cm}
\begin{center}
	\caption{Schematic structure of the higher gauge symmetries: the different shapes of the objects indicate different scaling dimensions (see below in the text) of canonical variables. The directions of arrows indicate how the variables are  mixed into others (or into themselves) by gauge transformations. 
The row in the middle represents conserved vectors, while the top row
represents the corresponding cyclic variables. 
}
\end{center}
\end{minipage}
\end{figure}

We assume as usual that no $\tau$-derivatives are allowed in the 
Hamiltonian in terms of the canonical variables. This 
essentially amounts to assuming that the equations of motion 
do not involve higher derivatives than the second 
derivative. 
Then the above Poincar\'{e} invariant tells us the 
following four Gauss constraints that are 
obtained by the variational principle with 
respect to the gauge fields.  
\begin{align}
	&\delta\boldsymbol{A}: \quad [\boldsymbol{P}_{\mu}, 
	\boldsymbol{X}^{\mu}]+\ldots \approx 0, \label{gauss1}\\
	&\delta\boldsymbol{B}:\quad 
	\hat{\boldsymbol{P}}_{\mu}X_{{\rm M}}^{\mu}\approx 0, 
	\label{gauss2}\\
	&\delta\boldsymbol{Z}:\quad 
	P_{\circ}^{\mu}\hat{\boldsymbol{X}}_{\mu}\approx 0, \label{gauss3}\\
	&\delta B:\quad P_{\circ}^{\mu}X_{{\rm M}\mu}
	\approx 0, \label{gauss4}
\end{align}
where the notation $\approx$ symbolizes that these 
constraints are regarded as {\it weak} equations 
before gauge fixing. This set of constrains is closed 
in the sense of Poisson bracket algebra and hence is 
of first-class. 
The first one with the ellipsis being the 
contribution from the fermionic part 
is the requirement of the SU($N$) invariance, 
that exists already in the original 
M(atrix) theory which is supposed to be 
obtained from our system by an appropriate gauge fixing 
in the light-like frame. The remaining three constraints 
are consequences from the higher gauge symmetries: 
the second corresponds to the shift symmetry originated 
from our definition of the 3-bracket, the third does 
to the $\delta_Y$ and the last does to the $\delta_w$. 

We make a natural assumption that the 
center-of-mass momentum $P_{\circ}^{\mu}$ and 
the M-variables $X_{{\rm M}}^{\mu}$ are 
conserved, $\frac{dP_{\circ}^{\mu}}{d\tau}=
\frac{dX_{{\rm M}}^{\mu}}{d\tau}=0$,
 as being associated 
with translation symmetries with respect to 
the center-of-mass coordinate 
$X_{\circ}^{\mu}$ and the M-component $P_{{\rm M}}^{\mu}$ of the 
canonical momentum variables, respectively; 
these are {\it cyclic} coordinates in our system. 
Since the $P_{\circ}^{\mu}$ is a time-like 
constant vector, we can assume by the 
last constraint that the $X_{{\rm M}}^{\mu}$ is 
a space-like constant vector. 
The third and the second constraints then imply
that 
the time-component of the traceless 
coordinate degrees of freedom $\hat{\boldsymbol{X}}^{\mu}$ 
and one of the space-components of the 
traceless momentum degrees of freedom $\hat{\boldsymbol{P}}^{\mu}$ become unphysical, respectively. 
This fits our requirement of reducibility 
of our system to the M(atrix) theory 
with SO(9) degrees of freedom alone for the 
traceless matrix canonical variables. 
We will see that the coordinate 
M-variable $X^{\mu}_{{\rm M}}$ determines
 the gauge-coupling constant, 
namely the expectation value of the dilaton 
in the sense of string theory. 

Now let us proceed to the full bosonic action of this 
system. The simplest possible gauge-invariant that is 
consistent with the M(atrix) theory is given as
\begin{align}
	A_{{\rm boson}}=
	\int d\tau \Bigl[&P_{{\rm M}\,\mu}
	\frac{dX_{{\rm M}}^{\mu}}{d\tau}+
	P_{\circ\,\mu}\frac{DX_{\circ}^{\mu}}{D\tau}+{\rm Tr}
	\Bigl(\hat{\boldsymbol{P}}_{\mu}\frac{D\hat{\boldsymbol{X}}^{\mu}}
	{D\tau}
	\Bigr) \nonumber \\
	&-\frac{e}{2N}P_{\circ}^2
	-\frac{e}{2}{\rm Tr}(\hat{\boldsymbol{P}}-
	P_{\circ}\boldsymbol{K})^2
	+\frac{e}{12}\langle[X^{\mu},X^{\nu},X^{\sigma}
	][X_{\mu},X_{\nu},X_{\sigma}]\rangle\Bigr], 
\end{align}
where the $\boldsymbol{K}$ is an auxiliary 
traceless matrix variable which is 
introduced for the purpose of recovering the $\delta_Y$ 
gauge symmetry for the quadratic kinetic 
term $\hat{\boldsymbol{P}}^2$ in the Hamiltonian 
with the transformation law $\delta_Y\boldsymbol{K}$. 
The situation is analogous to the Stueckelberg formalism 
of the well-known gauge invariant formulation of a massive 
Abelian vector field: after a gauge choice 
$\boldsymbol{K}=0$ for the gauge transformation $\delta_Y$, 
the equations of motion 
for $\boldsymbol{K}$ become the Gauss-type constraint,
\begin{align}
	P_{\circ}^{\mu}\hat{\boldsymbol{P}}_{\mu}=0,
	\label{gauss5}
\end{align}
which, together with \eqref{gauss3}, eliminates 
completely the time component of the 
canonical pair of the traceless matrix variables. 

The variation of the ein-bein $e$ gives the 
mass-shell condition 
\begin{align}
	P_{\circ}^2+{\cal M}^2_{{\rm boson}}\approx 0
\end{align}
where the effective mass-square is 
\begin{align}
	{\cal M}_{{\rm boson}}^2=N
	{\rm Tr}\hat{\boldsymbol{P}}^2
	-\frac{N}{6}\langle[X^{\mu},X^{\nu},X^{\sigma}
	][X_{\mu},X_{\nu},X_{\sigma}]\rangle
\end{align}
where we have assumed the gauge condition $\boldsymbol{K}=0$ 
and all of the Gauss-type constraints \eqref{gauss1}
$\sim$ \eqref{gauss4} and \eqref{gauss5}.
The second term, `potential energy', in this expression 
takes the form 
\begin{align}
	{\cal V}\equiv -\frac{N}{2}{\rm Tr}
	\Bigl(X_{{\rm M}}^2[\boldsymbol{X}^{\mu},
	\boldsymbol{X}^{\sigma}]
	[\boldsymbol{X}_{\nu},\boldsymbol{X}_{\sigma}]
	-2[X_{{\rm M}}\cdot \boldsymbol{X}, \boldsymbol{X}^{\nu}]
	[X_{{\rm M}}\cdot \boldsymbol{X}, \boldsymbol{X}_{\nu}]
	\Bigr),
\end{align}
in terms of the matrix notation. 

It is easy to check that the equations of motion 
are consistent with these constraints. 
It should be noted that the cyclic coordinates $X_{\circ}^{\mu}$ and 
$P_{{\rm M}}^{\mu}$ do not participate in the dynamics 
and, as such, are determined passively by other independent 
dynamical variables. For instance, 
the $P_{{\rm M}}^{\mu}$ obeys $\frac{DP_{{\rm M}}^{\mu}}{D\tau}
=\frac{\partial {\cal V}}{\partial X_{{\rm M}\mu}}$. 
As a matter of course, the center-of-mass momentum 
is fixed by the Hamiltonian constraint ({\it i.e} the 
mass-shell condition above), which in turn 
determines the center-of-mass coordinate 
through the equations of motion, $\frac{DX_{\circ}^{\mu}}
{D\tau}=\frac{e}{N}P^{\mu}_{\circ}$. 

One of the remarkable properties of the above action is that 
it is invariant under the following scale transformation:
\begin{align}
	&\boldsymbol{X}^{\mu}\rightarrow \lambda\boldsymbol{X}^{\mu}, \quad \boldsymbol{P}^{\mu}
	\rightarrow \lambda^{-1}\boldsymbol{P}^{\mu}, 
	\quad X_{{\rm M}}^{\mu}\rightarrow \lambda^{-3}X_{{\rm M}}^{\mu}, \quad P_{{\rm M}}^{\mu}
	\rightarrow \lambda^3 P_{{\rm M}}^{\mu}, \\
	&\boldsymbol{A}\rightarrow \lambda^{-2}\boldsymbol{A},
	\quad  \boldsymbol{B}\rightarrow \lambda^2\boldsymbol{B}, 
	\quad \boldsymbol{H}\rightarrow\boldsymbol{H}, 
	\quad \boldsymbol{L}\rightarrow\lambda^4\boldsymbol{L}, 
	\\
	&B_{\circ}\rightarrow \lambda^2B_{\circ}, 
	\quad \boldsymbol{Z}\rightarrow \lambda^{-2}
	\boldsymbol{Z}, \quad \boldsymbol{K}
	\rightarrow \boldsymbol{K}.
\end{align}
The emergence of such a scale symmetry is not surprising 
if we recall that the effective Yang-Mills gauge theory
 can also 
be regarded as being invariant under the 
``generalized conformal symmetry" which has been playing 
some useful roles in extending the AdS/CFT 
correspondence for ``non-dilatonic"  
branes such as D3-branes to the so-called ``dilatonic" 
D-branes in the sense of the GKPW relation. 
See {\it e.g.} \cite{sekiyone} and \cite{hnsy} 
and references therein. 
We now clearly see also that a role of 
the M-variable $X_{{\rm M}}^{\mu}$ as 
a conserved space-like vector is to set the length scale 
in our system by fixing its absolute value, namely, 
 the coupling constant for the potential ${\cal V}$ as 
we have already alluded to. 
Furthermore, the choice of its 
space-like direction 
in the ambient 11 dimensional space-time, 
together with the time-like direction of another 
conserved vector $P_{\circ}^{\mu}$, 
fixes a particular two-dimensional plane 
embedded in the 11 dimensions, 
to which the transverse SO(9) directions ($\mu
\rightarrow i=1,2,\ldots,9$) 
are orthogonal. We call this plane ``M-plane". 

Let us 
check explicitly how we can obtain 
the BFSS M(atrix) theory in the light-like gauge. 
The M-plane is foliated by the light-front 
coordinates $P_{\circ}^{\pm}\equiv P_{\circ}^{10}
\pm P_{\circ}^0, X_{{\rm M}}^{\pm}\equiv X_{{\rm M}}^{10}
\pm X_{{\rm M}}^0$, satisfying $P_{\circ}^+X_{{\rm M}}^-
+P_{\circ}^-X_{{\rm M}}^+\approx 0$. Since 
$X_{{\rm M}}^{\mu}$ and $P_{{\circ}}^{\mu}$ are space-like 
and time-like, respectively, we can assume 
that $X_{{\rm M}}^+$ and $P_{\circ}^+$ are not zero. Then, by the 
$\delta_L$ gauge transformation, we can choose the 
gauge such that 
$	\hat{\boldsymbol{X}}^+=0$, 
which allows us to eliminate the $\hat{\boldsymbol{X}}^-$ 
by the Gauss constraint \eqref{gauss3} that can now be regarded as 
a strong equation:
\begin{align}
	0=P_{\circ}^+\boldsymbol{X}^++P_{\circ}^-
	\hat{\boldsymbol{X}}^-=P_{\circ}^-
	\hat{\boldsymbol{X}}^-	\quad \Rightarrow \quad 
	\hat{\boldsymbol{X}}^-=0.
\end{align}
For the momentum variables, the first-order equations of motion 
\begin{align}
	\hat{\boldsymbol{P}}^{\pm}=
	\frac{1}{e}\frac{d\hat{\boldsymbol{X}}^{\pm}}{d\tau}
	+i[\boldsymbol{A},\hat{\boldsymbol{X}}^{\pm}]
	-\boldsymbol{B}X_{{\rm M}}^{\pm}
	\quad 
	\Rightarrow\quad -\boldsymbol{B}X_{{\rm M}}^{\pm}=0
\end{align}
allows us to conclude $\boldsymbol{B}=\hat{\boldsymbol{P}}^{\pm}=0$ 
by using the Gauss constraint \eqref{gauss2}. Thus the light-like 
components of the traceless dynamical matrix degrees of freedom 
are completely eliminated. The effective mass square then 
takes the form 
\begin{align}
	{\cal M}_{{\rm boson}}^2=N{\rm Tr}\Bigl(
	\hat{\boldsymbol{P}}_i^2-
	\frac{1}{2}X_{{\rm M}}^2
	[\boldsymbol{X}_i,\boldsymbol{X}_j]^2\Bigr).
\end{align}
Thus the conserved Lorentz invariant $X_{{\rm M}}^2$ 
determines the 11 dimensional gravitation length as
\begin{align}
	X_{{\rm M}}^2=\frac{1}{\ell_{11}^6}. 
\end{align}
The scaling symmetry is spontaneously broken 
by this identification. It seems natural to 
interpret this emergence of the fundamental scale 
of M-theory as a super-selection rule in the 
sense that we do not allow superposition of 
states with different values of the Lorentz invariant 
$X_{{\rm M}}^2$ in 
the Hilbert space after quantization. 

It is easy to check that 
the equations of motion for the center-of-mass variables 
are reduced to the usual ones $P_{\circ}^{\pm}
=N\frac{dX_{\circ}^{\pm}}{ds}$ under the 
gauge choice $B_{\circ}=0$ for the $\delta_w$-gauge 
transformation, where we defined the reparametrization 
invariant time parameter $s$ by 
$ds=ed\tau$. To summarize all, the effective action 
for the physical transverse variables can be expressed as 
\begin{align}
	A_{\rm light-front}=
	\int dx^+\frac{1}{2R}{\rm Tr}
	\Bigl[\Bigl(\frac{D\hat{\boldsymbol{X}}_i}{Dx^+}\Bigr)^2
	+\frac{R^2}{2\ell_{11}^6}[\boldsymbol{X}_i,\boldsymbol{X}_j]^2
	\Bigr]. 
\end{align}
Here, we have redefined the time parameter by 
$s=2Nx^+/P_{\circ}^+$ (or equivalently $X_{\circ}^+=2x^+
$). This reproduces the identification of the continuous 
parameter $R$ by \eqref{Pplus} in subsection 4.1. 

Finally, we would like to add a remark on some 
peculiar nature of the higher gauge symmetry. 
The fact that we can eliminate both of the traceless 
parts, $\hat{\boldsymbol{X}}^{\pm}$, of the matrix 
degrees of freedom means that the space-time 
directions corresponding to the M-plane are 
locally unobservable with respect to the 
dynamics of M-theory partons. In contrast to this, 
another light-like component $X^-$ of 
a single string (or of single membrane) is non-vanishing 
in the light-front 
gauge $\partial_{\sigma}X^+=0$:
 it is expressed in 
terms of the transverse components and 
behaves  
as a {\it passive} variable that does not participate 
in the dynamics. In our case, if $X_{{\rm M}}^+
\hat{\boldsymbol{X}}^-$ 
in the potential ${\cal V}$ were not eliminated, 
we would have a term 
$-\bigl(X_{{\rm M}}^+X^-_{ab}(x_a^i-x_b^i)\bigr)^2$ 
giving non-zero potential of wrong sign for 
purely diagonal configurations of the 
transverse directions. The absence of this term 
is consistent of a remarkable aspect 
of general-relativistic interactions of 
M-theory partons that the bunldes of parallel 
trajectories of partons are exact classical solutions. 
It should be remembered here that 
the parallel pencil-like trajectories 
of massless particles are non-interacting to 
each other in the classical 
11 dimensional general relativity. This 
corresponds to the well known property that 
for the metric of the form 
\begin{align}
	ds^2=dx^{\mu}dx_{\mu}+h_{--}(dx^-)^2
\end{align}
with the coordinate condition $\partial_+h_{--}=0$, 
the vacuum Einstein equation reduces to the linear 
Laplace equation $\partial_i^2h_{--}=0$ 
in the transverse space. 
These properties make possible the interpretation 
of states with higher quantized momenta $P_{\circ}^+$ 
as composite states consisting of constituent 
states with unit momentum $1/R$ along the 
compactified spatial direction. 

There are many remaining problems left to the 
future. Most importantly, although 11-dimensional Lorentz 
covariance is realized at least kinematically, 
it is not clear whether this reformulation of 
the M(atrix) theory may lead to further 
insight on the nature of the M-theory conjecture, especially 
its non-perturbative dynamical aspects 
of string/M theory. I can only hope that 
the present discussion would be an intermediate 
step toward such a goal that has not been 
attained for more than 
two decades since its first inception occured
during various explosive developments in the 1990s. 
In connection with this, the generalization 
of the present formalism to matrix-string theory 
seems to be relevant. For example, we may try 
to make covariant the procedure adopted in \cite{sekiyone2} 
(to which I would like to refers readers  
for a bibliography) 
for deriving matrix-string theory 
directly from the classical theory of (super) 
membranes. The emergence of the Nambu-type brackets 
may also be useful in extending 
the system to various non-trivial
 backgrounds (see, {\it e.g.}  \cite{sheikh}).

\section{Generalized Hamilton-Jacobi formalism 
of Nambu mechanics}
We now return to Nambu's original 
motivation of devising a new possible canonical 
formulation of dynamical development 
and its quantization. Concerning quantization, 
there have been a number of 
further discussions continuing and improving 
Nambu's 
attempt in various directions. 
However, it seems to be 
a remarkable 
fact that there has been no serious discussion 
in the spirit of {\it wave}-mechanical 
quantization, until the attempt \cite{yonegHJ} of the 
present author in which 
a Hamiltonian-Jacobi like reformulation of 
the Nambu mechanics was proposed as 
a plausible 
prerequisite for quantization. That formulation then 
suggested a natural approach to quantum theory. 
As in the previous section, the purpose of this section 
is to concentrate on my motivation toward 
such a direction and explain 
the basic 
ideas of this approach, leaving the details 
to this reference. 

In fact, a naive thinking along the 
traditional way of formulating 
the usual Hamilton-Jacobi (HJ) formalism in the 
ordinary analytical dynamics suggests   
the following obstacles against such a direction: 
\begin{enumerate}
	\item[(i)] The canonical triplet $\xi^i$ ($i=1,2,3$) does not lend natural decomposition of the phase space into pairs of generalized coordinates and momenta.
	\item[(ii)] There is no known explicit 
	formulation of finite canonical transformation, 
	as opposed to the infinitesimal one. 
	This is closely related to what we have 
	discussed in subsection 2.3. Remember that the 
	usual text-book formulation of the HJ  
	formalism rests upon the generating function 
	for finite canonical transformation. 
	\item[(iii)] There is no action integral defined for each one-dimensional trajectory in the phase space, as opposed to a known action \cite{takh} function for a {\it continuous family}, forming a two-dimensional surface, of such trajectories.
\end{enumerate}
In connection with (iii), it should 
be emphasized again that the action principle 
of \cite{takh} suffers from an infinite redundancy 
from the viewpoint of dynamics since 
a continuous family of trajectories has 
only kinematical meaning: 
the string-like world surface as the continuous 
collection of one-dimensional trajectories has nothing to do with the physical force or 
tension along the `$\sigma$' direction. 
This feature perhaps correlates 
with the difficulty (ii). 

A similar infinite redundancy also appears when 
we try to quantize the local 
dynamics of Yang-Mills field 
in terms of the set of non-local 
gauge invariant Wilson 
loops. In that case, however, we 
can express the local dynamics of quantum 
Yang-Mills 
fields in the form of a non-local field theory
by deriving appropriate set of constraints 
for the string fields corresponding to 
the Wilson loops, such that the result 
gives a theory of string fields 
which is in principle equivalent with the 
usual formulation in terms of the original 
local fields. For an explicit construction of such a dynamical system 
of non-local string fields that is guaranteed to be 
equivalent with the usual local formulation 
in the context of lattice gauge theory, see my 
early work 
\cite{yonelattice} and references therein. 
In this case, there is also an important 
physical motivation for such a direction, 
since in the long-distance strong-
coupling regime the string-like objects of colored 
electric fluxes become the dominant 
physical excitations. Furthermore, 
 that would also 
be relevant for  
weak-coupling regime in the large $N$ limit 
from the viewpoint of the planar expansion of Feynman 
graphs.

By constrast, in the case of the Nambu mechanics, 
we have to keep in mind that no 
definite and universally acceptable
 quantum formulation in terms 
of the local dynamical variables has 
been established. There do not seem sound 
physical motivations analogous to the 
flux strings or the 
large $N$ limit, either. Indeed, what we have 
discussed in section 4 is the application of the 
Nambu equations of motion 
{\it not} for realizing the dynamics of membranes, but 
only for a clue toward higher symmetries within the 
framework of usual Hamiltonian dynamics. 
In my viewpoint, it is important to discriminate 
symmetry and dynamics. In the present 
section, the main focus is on the latter, though 
the former will also play en essential role. 
 
Historically, there have been several attempts at 
generalizing HJ formalism to systems 
with higher symplectic forms. For a convenient review, 
the reader is recommended to consult \cite{kastrup}. 
Indeed, any field theory, including world-volume 
theories of extended objects can formally 
have a higher symplectic form whose rank coincides 
with the dimensions of the object, as we dealt 
with the action \eqref{momentuaction}. However, all those 
works have concerned only about the various 
cases with a single Hamiltonian. As a notable example, 
we can mention Nambu's attempt \cite{nambuHJstring} 
toward a novel HJ-like 
formulation of strings which appeared in 1980. 
That seems to be in fact equivalent with the earlier 
De Donder-Weyl formalism as reviewed in \cite{kastrup}. 

\subsection{Preliminaries}
Now we can proceed to our main 
theme of this section. Fortunately, there is an interesting 
reformulation of the HJ formalism that does not 
presuppose any knowledge of an action functional 
nor of a concrete form for finite canonical transformations, 
starting from scratch with only 
the equations of motion. 
This was suggested by Einstein hundred years ago 
in an almost unknown short paper \cite{ein1917} 
where he emphasized that his method was 
``free of surprising tricks of trades". 
His arguments arose from his 
attempt \cite{ein2} at generalizing the Sommerfeld-Epstein 
quantization to non-separable cases and 
giving a coordinate-independent formulation of 
semi-classical quantum theory. 
In this subsection, a derivation of the standard HJ 
formalism {\it \`{a} la} Einstein will be reviewed first 
for the usual Hamiltonian formalism. 
After this preparation, 
generalization to the Nambu mechanics will be given 
in the next subsection. 

The basic idea is that the trajectories 
$q^i=q^i(t), p_i=p_i(t)$ 
obeying the Hamilton equations of motion in the usual 
phase space
\begin{align}
	\frac{dq^i}{dt}=\frac{\partial H}{\partial p_i}, \quad 
	\frac{dp_i}{dt}=-\frac{\partial H}{\partial q^i}
\end{align}
are re-formulated to describe the time developments 
of the momentum $p_i$ as the fields $p_i=p_i(q,t)$ 
defined on the configuration space of the 
the coordinates $q^i$'s. Namely, using the modern terminology, 
the phase space is interpreted as a 
fiber bundle (called the `cotangent bundle') 
over the configuration space, $p_i$'s and 
$q_i$'s being the parameters describing   
the fiber and base space, respectively. 
The momentum part of the Hamilton 
equations of motion is then rewritten, 
using the coordinate part, as
\begin{align}
	\frac{dp_i}{dt}=
	\frac{\partial p_i}{\partial t}+
	\frac{\partial H}{\partial p_j}\frac{\partial p_i}{\partial q^j}=-\frac{\partial H}{\partial q^i}.
	\nonumber
\end{align}
The second equality 
\begin{align}
	\frac{\partial p_i}{\partial t}+
	\frac{\partial H}{\partial p_j}\frac{\partial p_i}{\partial q^j}+\frac{\partial H}{\partial q^i}=0\end{align}
can be interpreted  
as the partial differential equations of 
first order for the vector fields 
$p_i(q,t)$. Let us call this type of 
equations an ``Euler-Einstein" (EE) equation, 
since it is analogous to Euler equations 
of motion in fluid mechanics. 
In order to avoid possible confusions, 
we introduce the notation
\begin{align}
	\bar{H}=\bar{H}(q,t)=H\bigl(p(q,t),q\bigr), 
\end{align}
in terms of which the EE equation is 
$\frac{\partial p_i}{\partial t}+\frac{\partial \bar{H}}{\partial q_i}=0$.
The next step is to require that the flow 
described by the EE equation has 
no vorticity 
and hence can be expressed as a gradient flow 
associated with a scalar potential field $J=J(q,t)$: 
\begin{align}
	\frac{\partial p_j}{\partial q_i}-
	\frac{\partial p_i}{\partial q_j}=0 
	\quad 
	\Rightarrow \quad p_i=\frac{\partial J}{\partial q^i}. 
\end{align}
This requirement replaces the demand for 
the existence of the generating function 
arising from the action in the text-book formulation 
of the HJ formalism. Einstein's motivation for 
this requirement is that the action integral 
$\int p_idq^i$ along a curve 
in the configuration space should take the same value 
for all deformations that can 
be continuously connected to each other, at least locally, 
 under the 
condition of fixed end points. That would assign 
 a special invariant meaning to the integral $\oint p_idq^i$ which is subjected to 
 the Sommerfeld-Epstein 
condition for quantization. 
Then, the EE equation takes the form 
\[
\frac{\partial}{\partial q^i}\Bigl(
\frac{\partial J}{\partial t}+\bar{H}\Bigr)=0
\quad 
\Rightarrow \quad \frac{\partial J}{\partial t}+\bar{H}=f(t)
\]
where $f$ is an arbitrary function of time only. But 
we can eliminate this arbitrariness by redefining 
$J$ with $\partial_tS\equiv \partial_tJ-f$. 
Thus we arrive at the HJ equation
\begin{align}
	\frac{\partial S}{\partial t}+\bar{H}=0, 
	\quad p_i=\frac{\partial S}{\partial q^i}. 
\end{align}
It seems fairly obvious that this process 
does not depend on the dimensionality of 
the phase space and the configuration space, 
nor on the existence of 
the action integral, at least explicitly. 
The dimensions of the fiber and base space 
can have {\it different} dimensions.

It is a standard matter to
 reverse this process, obtaining the Hamilton 
equations of motion for trajectories $\bigl(q^i(t),p_i(t)\bigr)$ 
starting from the HJ equation. 
For the sake of reminder, let us recall its 
essence here. 
First consider the momentum vector fields obtained from a 
solution of the HJ equation. They satisfy 
\begin{align}
	\frac{dp_i}{dt}=\frac{\partial p_i}{\partial t}+\frac{\partial p_i}{\partial q^j}
	\frac{dq^j}{dt}=
	\frac{\partial^2 S}{\partial q^i\partial t}+\frac{\partial^2S}{\partial q^j\partial q^i}
	\frac{dq^j}{dt}
	=-\frac{\partial H}{\partial q^i}-\frac{\partial H}{\partial p_j}
	\frac{\partial^2 S}{\partial q^j\partial q^i}
	+\frac{\partial^2S}{\partial q^j\partial q^i}
	\frac{dq^j}{dt}=-\frac{\partial H}{\partial q^i}
	\nonumber
\end{align}
provided we know that the $q^i$'s satisfy 
the coordinate part of the Hamilton equations of motion. 
Therefore our problem is reduced to deriving the latter. 
Given the `complete' solution of the HJ equation, 
this is achieved by imposing the so-called Jacobi condition. Remember 
that a `complete' solution is a solution 
of the HJ equation with $n$ 
independent integration 
constants $Q_i$ when the dimensions of the configuration space is  $n$. 
Then the $S$ can be regarded as a function $S=S(q, Q;t)$ 
of the $2n$ independent variables $(q_i,Q_i)$ and $t$. In the usual text-book derivation 
of the HJ equation, $q^i$'s and $Q^i$s 
 actually arise as the set of initial and final 
coordinates, respectively, for general trajectories 
satisfying the equations of motion.  
The Jacobi condition is the following restrictions 
imposed upon a complete solution 
\begin{align}
	\frac{\partial S}{\partial Q^i}=-P_i,
	\label{jacobi}
\end{align}
with an additional set of $n$ integration constants $P_i$. 
Under the solvability condition 
\begin{align}
	{\rm det}\Bigl(
	\frac{\partial^2 S}{\partial Q^i\partial q^j}
	\Bigr)\ne 0, 
	\label{sol}
\end{align}
they implicitly determine 
the coordinates $q^i$'s as the functions of time, 
satisfying the coordinate part of the Hamilton 
equations of motion: 
by performing a total differentiation 
of the Jacobi condition with respect to time $t$, 
we obtain
\begin{align}
	0=\frac{\partial^2 S}{\partial Q^i\partial t}+
	\frac{\partial^2 S}{\partial Q^i\partial q^j}
	\frac{dq^j}{dt}=-\frac{\partial \bar{H}}{\partial Q^i}
	+\frac{\partial^2 S}{\partial Q^i\partial q^j}
	\frac{dq^j}{dt}=
	-\frac{\partial H}{\partial p_j}\frac{\partial^2S}{\partial q^j\partial Q_i}+\frac{\partial^2 S}{\partial Q^i\partial q^j}
	\frac{dq^j}{dt},\nonumber
\end{align}
which indeed leads to the desired equations 
under \eqref{sol}. 
The Jacobi condition is the clue to the quantization. 

In the language of differential forms, 
the above process of arriving at the HJ equation 
is summarized as follows: 
define a closed (and exact) 2-form in the phase space 
adjoined by a time variable, 
\begin{align}
	\omega^{(2)}=dp_i\wedge dq^i-dH\wedge dt=d\omega^{(1)}, 
	\quad 
	\omega^{(1)}=p_idq^i-Hdt. 
\end{align}
Then, the demand that the 2-form 
$\omega^{(2)}$ vanishes when it is evaluated 
after making projection to the configuration space
by assuming $p_i=p_i(q,t)$ reproduces the 
EE equation.
\begin{align}
	\bar{\omega}^{(2)}\equiv 
	\omega^{(2)}|_{(q,t)}=\frac{1}{2}
	\Bigl(\frac{\partial p_j}{\partial q^i}-
	\frac{\partial p_i}{\partial q^j}\Bigr)
	dq^i\wedge dq^j-
	\Bigl(\frac{\partial p_i}{\partial t}+
	\frac{\partial \bar{H}}{\partial q^i}\Bigr)dq^i\wedge dt=0. 
\end{align}
Then, the 1-form $\omega^{(1)}$ in the 
phase space must be an exact form
\begin{align}
	\bar{\omega}^{(1)}\equiv \omega^{(1)}|_{(q,t)}
	=d\bar{\omega}^{(0)}=\frac{\partial S}{\partial q^i}dq^i+\frac{\partial S}{
	\partial t}dt
\end{align}
in terms of a scalar function $\bar{\omega}^{(0)}\equiv S(q,t)$ after the projection. 
This is nothing but the HJ equation. 
As a matter of fact, this is the well known modern mathematical 
form of the HJ formalism \cite{arnold} (see also \cite{kastrup}), 
which is usually explained on the basis of the 
existence of the action integral, since the 
function $S$ as the generator of finite 
canonical transformation describing the 
equations of motion is originated from the action. 


\subsection{Generalized HJ formalism for Nambu mechanics}
For generalizing the HJ formalism to 
the Nambu mechanics along 
the foregoing procedure, it is convenient to set three steps 
as follows.
\begin{enumerate}
 \item[(I)] The EE equations from the 
 equations of motion: the phase space $(\xi^1, \xi^2,\xi^3)$
 is decomposed into fiber and base space. 
 \item[(II)] The HJ-like equations from the EE equations: 
we have to find appropriate vanishing condition under the 
projection to the base space, in the sense explained above. 
\item[(III)] The equations of motion from the HJ equation: 
we have to find appropriate Jacobi-like condition.  
\end{enumerate} 
The steps (I) and (II) are relatively straightforward, 
but the last step (III) that is most crucial 
for quantization will turn out to be 
rather difficult, since at least naively the 
solvability condition \eqref{sol} associated with 
\eqref{jacobi} seems to require the same 
dimensions for the configuration space 
and momentum space as the fiber directions. 

There are two possibilities 
for the decomposition, 
\begin{align}
&(1/2) \mbox{decomposition} : (\xi^1,\xi^2,\xi^3)
\rightarrow \bigl(\xi^1,\xi^2, \xi^3(\xi^1,\xi^2,t)\bigr), \nonumber \\
&(2/1) \mbox{decomposition} : (\xi^1,\xi^2,\xi^3)
\rightarrow \bigl(\xi^1, \xi^2(\xi^1,t), 
\xi^3(\xi^1,t)\bigr). \nonumber
\end{align}
In this review, only the first case will be 
treated in some detail. For the second case, the final 
results will be described briefly. 

\subsubsection{Step (I)}

We define
\begin{align}
	&\bar{H}=\bar{H}(\xi^1,\xi^2,t)=H\bigl(\xi^1, \xi^2, 
	\xi^3(\xi^1,\xi^2,t)\bigr),\\
	&\bar{G}=\bar{G}(\xi^1,\xi^2,t)=G\bigl(\xi^1, \xi^2, 
	\xi^3(\xi^1,\xi^2,t)\bigr).
\end{align}
The EE equation for the field $\xi^3=\xi^3(\xi^1,\xi^2,t)$ 
is derived by the same method as in the usual 
Hamilton equations of motion. 
For notational brevity, the partial derivatives 
with respect to the base space coordinates and 
time 
will be abbreviated as $\partial_i$ and $\partial_t$. 
Using the Jacobian form \eqref{Neqjacobian} of the Nambu equation, we obtain
\begin{align}
	\frac{\partial(H,G)}{\partial(\xi^1,\xi^2)}
	=\partial_t\xi^3+
	\partial_1\xi^3
	\frac{\partial(H,G)}{\partial(\xi^2,\xi^3)}
	+\partial_2\xi^3\frac{\partial(H,G)}{\partial(\xi^3,\xi^1)}.
	\label{xi3flow} 
\end{align}
On the other hand, the Jacobians are in terms 
of $\bar{H}$ and $\bar{G}$ are rewritten, using 
$\partial_i\bar{H}=\partial_iH+\partial_3H\partial_i\xi^3, 
\partial_i\bar{G}=\partial_iG+\partial_3G\partial_i\xi^3$as 
\begin{align}
	&\frac{\partial(H,G)}{\partial(\xi^2,\xi^3)}
	=\partial_2\bar{H}\partial_3G-\partial_2\bar{G}
	\partial_3H, \quad 
	\frac{\partial(H,G)}{\partial(\xi^3,\xi^1)}
	=\partial_3H\partial_1\bar{G}-\partial_3G\partial_1\bar{H},\nonumber\\
	&\frac{\partial(H,G)}{\partial(\xi^1,\xi^2)}=
	\partial_1\bar{H}\partial_2\bar{G}-
	\partial_1\bar{G}\partial_2\bar{H}
	-\partial_1\xi^3(\partial_3H\partial_2\bar{G}-
	\partial_3G\partial_2\bar{H})-
	\partial_2\xi^3(\partial_1\bar{H}\partial_3G-
	\partial_1\bar{G}\partial_3H)\nonumber. 
\end{align} 
By substituting these expressions to \eqref{xi3flow}, 
we finally obtain
\begin{align}
	\partial_t\xi^3=\partial_1(\bar{H}\partial_2\bar{G})
	-\partial_2(\bar{H}\partial_1\bar{G}), 
	\label{EE1}
\end{align}
which is indeed the desired EE equation, since the 
r.h.side is a known algebraic function 
of $\xi^1,\xi^2,\xi^3(\xi^1,\xi^2,t)$ and 
the derivatives $\partial_i\xi^3(\xi^1,\xi^2,t)$. 
This finishes the step (I). 

\subsubsection{Step (II)}
The result of the step (I) naturally suggests  
the answer to step (II). \eqref{EE1} shows 
that the field $\xi^3$ is expressed as the vorticity 
of a two-component vector field $(S_1,S_2)$ as 
\begin{align}
	\xi^3\equiv \epsilon^{3ij}\partial_iS_j=\partial_1S_2-
	\partial_2S_1. 
\end{align}
The EE equation then takes the form of a set of 
paired partial differential equations 
for $S_i$:\footnote{Please note that there are
trivial typos in eq. (25) of \cite{yonegHJ}.}
\begin{align}
	\partial_tS_i=\bar{H}\partial_i\bar{G}+\partial_iS_0,
	\label{gHJ1}
\end{align}
where $S_0$ is an arbitrary scalar field
and 
\begin{align}\bar{H}=H(\xi^1,\xi^2,\partial_1S_2-\partial_2S_1), 
\quad \bar{G}=G(\xi^1,\xi^2,\partial_1S_2-\partial_2S_1).
\end{align}

In this formalism, the N-gauge 
symmetry is effectively absorbed as a degree of 
freedom of making a redefinition
$S_0\rightarrow S_0-\Lambda$, keeping the 
r.h.side of \eqref{gHJ1} in the 
transformation of the pair of 
Hamiltonians $(H,G)$ into a new pair 
$(H',G')$.   
Independently of the 
N-gauge symmetry, this system is 
characterized by a new gauge symmetry of its own, 
\begin{align}
	S_{\mu}\rightarrow S_{\mu}+\partial_{\mu}\lambda
\end{align}
where $S_{\mu}$ $(\mu=1,2,0)$ is the three-dimensional 
vector field $(S_1,S_2,S_0)$ on the base 
`space-time' $\xi^{\mu}=(\xi^1,\xi^2,t)$. 
We call this extended gauge symmetry ``S"-gauge 
symmetry. The S-gauge symmetry is nothing to do 
with the gauge symmetry associated with the gauge 
field $A_i$ of \eqref{gaugefield}. Note that 
the latter is now connected to the 
field strength $F^S_{i0}\equiv 
\partial_iS_0-\partial_tS_i$ which is 
{\it invariant} under the S-gauge transformations. 
As a matter of course, this gauge symmetry suggests the 
Wilson lines $e^{\oint S_{\mu}d\xi^{\mu}}$ 
as a natural set of gauge-invariant observables. 
But, we have already emphasized that 
use of them as basic dynamical variables would 
annoy us with too much and unnecessary redundancy. 

The above result can be recast in the language of 
differential forms as follows. 
Obviously, we have already a natural 1-form 
on the base space-time
\begin{align}
	\bar{\Omega}^{(1)}\equiv S_{\mu}d\xi^{\mu}.
\end{align}
The generalized HJ equation is then 
expressed by the condition
\begin{align}
	\bar{\Omega}^{(2)}\equiv d\bar{\Omega}^{(1)}
	&=(\partial_1S_2-\partial_2S_1)d\xi^1\wedge d\xi^2
	+(\partial_iS_0-\partial_0S_i)d\xi^i
	\wedge dt \nonumber \\
	&=\xi^3d\xi^1\wedge d\xi^2-\bar{H}
	\partial_i\bar{G}d\xi^i\wedge dt.
\end{align}
The EE equation \eqref{EE1} is nothing but the 
vanishing condition for the 3-form 
$\bar{\Omega}^{(3)}=d\bar{\Omega}^{(2)}$. 
\begin{align}
	0=\bar{\Omega}^{(3)}=\partial_0\xi^3d\xi^1
	\wedge d\xi^2\wedge dt-
	(\partial_1\bar{H}\partial_2\bar{G}-\partial_2\bar{H}
	\partial_1\bar{G})d\xi^1\wedge d\xi^2 \wedge dt. 
\end{align}
Unlike the standard HJ formalism, the fiber is 
not directly related to the tangent planes of the base 
space. Instead, the one-dimensional fiber parametrized 
by $\xi^3$ is the measure of 
vorticity in the base space. We may call our 
3 dimensional phase space a ``vorticity bundle". 

We can also rephrase the structure of 
the vorticity bundle before making the 
Einstein projection. We are led to the 
2-form on the total 4-dimensional space-time 
$(\xi^1,\xi^2,\xi^3, t)$ 
\begin{align}
	\Omega^{(2)}\equiv \xi^3d\xi^1\wedge d\xi^2
	-HdG\wedge dt
\end{align}
and the closed and exact 3-form 
\begin{align}
	\Omega^{(3)}\equiv d\Omega^{(2)}=
	d\xi^1\wedge d\xi^2\wedge d\xi^3
	-dH\wedge dG \wedge dt, 
\end{align}
such that they reduce to $\bar{\Omega}^{(2)}, 
\bar{\Omega}^{(3)}$ after the Einstein projection. 
The 2-form $\Omega^{(2)}$ coincides with the one 
that was used for defining the action integral \cite{takh} 
for 1-dimensional family of the Nambu-flow 
trajectories, as previously mentioned in subsection 3.1. 
That action principle is essentially equivalent to the 
the null condition for the 3-form $\Omega^{(3)}$, 
\begin{align}
	i_{L}(\Omega^{(3)})=0 \quad 
	\mbox{with} \quad 
	L\equiv X^i\partial_i
	+\partial_t,\quad 
	X^i\equiv \epsilon^{ijk}\partial_jH\partial_kG
	\label{null}
\end{align}
where the symbol $i_L$ is interior multiplication 
with respect to the vector field operator $L$ 
acting on the differential form defined on 
the 4-dimensional space-time. This is the 
analog of the similar property of $\omega^{(2)}$ 
of the ordinary Hamilton dynamics where 
the Hamiltonian flow with 
operator $V\equiv \epsilon^{ij}\partial_jH\partial_i+\partial_t$ 
satisfies the null condition $i_{V}(\omega^{(2)})=0$.\footnote{Because of the difference of the ranks 
between $\omega^{(2)}$ and $\Omega^{(3)}$,  
the action principle along one-dimensional trajectories 
is impossible for the Nambu mechanics. 
} 
Thus, the structure of the generalized HJ equation 
associated with 
the EE equation  
is precisely parallel to that in the standard 
Hamiltonian dynamics, provided the dimensions 
of the corresponding differential forms are 
up-graded by one: 
$(\bar{\omega}^{(1)},\bar{\omega}^{(0)}) 
\rightarrow (\bar{\Omega}^{(2)}, 
\bar{\Omega}^{(1)}), 
(\omega^{(2)},\omega^{(1)})
\rightarrow (\Omega^{(3)},\Omega^{(2)})$. 

\subsubsection{Step (III)}
A prerequisite for this step is that, 
given the trajectory $(\xi^1(t),\xi^2(t))$ 
satisfying the Nambu equations of motion, 
$\xi^3$ determined by the EE equation automatically 
satisfy the last of the Nambu equation. 
As in the ordinary HJ formalism, this is easily 
checked by reversing the process of obtaining 
the EE equation from the Nambu equations. 
The real problem in step (III) is to 
show how the first and the second components (the `coordinate' part)  
of the Nambu equations of motion are 
derived from the generalized HJ equations \eqref{gHJ1} 
by imposing some analog of the Jacobi condition. 

Here a fundamental difficulty arises when 
we try to follow the usual procedure. 
In the standard case, the rationale for the 
Jacobi condition \eqref{jacobi} is that we can unfold 
the dynamics to the vanishing Hamiltonian 
by finite canonical transformation whose 
generating function $S$ satisfies, after projection, 
\begin{align}
	p_idq^i-\bar{H}dt=dS+P_idQ^i. 
\end{align}
Namely, the $S$ can be treated as a function of $(q^i,
Q^i)$ obeying 
\begin{align}
	p_i=\frac{\partial S}{\partial q^i}, 
	\quad 
	P_i=-\frac{\partial S}{\partial Q^i}.
\end{align}
Since the corresponding form in the case of 
the Nambu mechanics is $\bar{\Omega}^{(2)}$, one would 
expect a natural extension to be something like
\begin{align}
	\xi^3d\xi^1\wedge d\xi^2-\bar{H}d\bar{G}
	\wedge dt=d\Sigma^{(1)} + Q_3dQ_1\wedge dQ_2
	\nonumber
\end{align}
for an unfolding transformation to the vanishing 
Hamiltonian with three integration constants denoted 
by $(Q_1,Q_2,Q_3)$ and a possible generating 1-form 
$\Sigma^{(1)}$. But this amounts to treating 
the generating 1-form $\Sigma^{(1)}$ as a function of 
four variables $(\xi^1,\xi^2; Q_1,Q_2)$, 
which is obviously wrong since we have 
only three independent canonical variables. 
For a more detailed discussion on this 
problem, readers are referred to 
the original paper \cite{yonegHJ}. 
In the last reference, we have suggested 
two ways to resolve this 
difficulty. In the present review, 
only the first one which is somewhat restricted 
but simpler than the second will be 
treated to some extent for the 
purpose of showing an attitude 
to the present problem concretely. 

The first resolution is 
to simplify the generalized HJ equation 
by assuming an axial gauge condition with respect to 
the N-gauge symmetry (or by an appropriate 
choice of canonical coordinates) for 
the choice of the set of Hamiltonians,
\begin{align}
	\partial_3G=0 
	\quad 
	\rightarrow \bar{G}=G(\xi^1,\xi^2), 
\end{align}
and to go to the $S_0=0$ gauge 
with respect to the S-gauge symmetry. 
It may not be always justified to make this 
simplification. However, 
as we will see later explicitly, this is possible 
in an important class of systems including 
the case of the Euler top \eqref{Eulertop1} 
mentioned in section 2. 
Then, we first set 
\begin{align}
	S_i=-S\partial_iG+g_i \nonumber
\end{align}
where the single function $S$ is defined by a 
solution to the ordinary-looking HJ equation
\begin{align}
	\partial_tS=-\bar{H}, 
	\label{axialgHJ}
\end{align}
by which we have
\begin{align}
	\partial_tg_i=0, 
	\quad \xi^3=\partial_1G\partial_2S-\partial_2G\partial_1S+\partial_1g_2-
	\partial_2g_1.
	\nonumber
\end{align}
Using the residual time-indenpendent 
$S$-gauge transformation, we can set $g_i=\epsilon^{ij}
\partial_jg$ with a time-independent 
scalar function $g$. Then, it is possible to 
absorb the $g$ into $S$ by making a redefinition 
$S\rightarrow S+F$ where $F$ is defined by
\begin{align}
	F\partial_iG=\epsilon^{ij}\partial_jg, 
\end{align}
which is indeed solved as 
\begin{align}
	d_GF=\partial_i\partial^ig, \quad 
	d_G\equiv \epsilon^{ij}\partial_iG\partial_j
\end{align}
by assuming at least one $\partial_iG$ is not zero. 
Therefore, without losing generality within
 the present gauge condition, 
we can set 
\begin{align}
	\xi^3=d_GS. 
	\label{xi3vortex}
\end{align}

The generalized HJ equation is now reduced to 
\eqref{axialgHJ} adjoined with \eqref{xi3vortex}. 
This is almost the usual form except for the 
difference that there is only one momentum 
defined by the latter equation (in spite 
of the two-dimensional configuration space $(\xi^1,
\xi^2)$), which is still 
a sort of vorticity composed by an outer product 
of two vectors $\partial_iG$ and $\partial_iS$, 
instead of the gradient of the usual HJ.  
Step (III) can now be achieved. 

Suppose we have a complete 
solution to \eqref{axialgHJ} with an  
integration constant $Q_1$. 
Since the value of $G$ is preserved 
against the only differential operation involved 
in \eqref{axialgHJ} due to 
$d_GG=0$, we can assume the value of $G=\bar{G}$ 
as the second integration constant $Q_2$. 
As the Jacobi condition, we 
impose 
\begin{align}
	\frac{\partial S}{\partial Q_1}=Q_3, 
	\label{jacobin1}
\end{align}
by introducing the third integration constant $Q_3$. 
These two conditions implicitly determine 
$\xi^1, \xi^2$ as functions of time, 
$\xi^i(t; Q_1,Q_2,Q_3)$, under the condition 
that \eqref{jacobin1} is not invariant against $d_G$, 
namely, 
\begin{align}
	\frac{\partial^2S}{\partial\xi^1\partial Q_1}\partial_2G-
	\frac{\partial^2S}{\partial\xi^2\partial_1Q}
	\partial_1G
	\ne 0.
	\label{soln1}
\end{align}
It is also possible to derive the condition 
\eqref{jacobin1} by considering the 
canonical transformation under the condition 
$\partial_3G=0$, as discussed in \cite{yonegHJ}. 

Let us confirm how the above Jacobi condition 
leads to the Nambu equations of motion 
for $\xi^1, \xi^2$. Taking a total 
time derivative of \eqref{jacobin1} and 
$G=Q_2$ yields 
\begin{align}
	\frac{\partial^2S}{\partial Q_1\partial t}
	+\frac{\partial^2S}{\partial \xi^1\partial Q_1}
	\frac{d\xi^1}{dt}+\frac{\partial^2S}{\partial\xi^2
	\partial Q_1}\frac{d\xi^2}{dt}&=0, \\
	\partial_1G\frac{d\xi^1}{dt}+\partial_2G
	\frac{d\xi^2}{dt}=0. 
\end{align}
Using \eqref{axialgHJ}, the first term of the first equation 
is rewritten as 
\begin{align}
	\frac{\partial^2S}{\partial Q_1\partial t}
	=-\frac{\partial H}{\partial \xi^3}
	\Bigl(\partial_1G\frac{\partial^2 S}{\partial \xi^2
	\partial Q_1}-
	\partial_2G\frac{\partial^2S}{\partial \xi^1
	\partial Q_1}\Bigr).
	\end{align}
Then, using the second equation, we can conclude, due to \eqref{soln1}, 
\begin{align}
	\frac{d\xi^1}{dt}=-\partial_3H\partial_2G, 
	\quad \frac{d\xi^2}{dt}=\partial_3H\partial_1G, 
\end{align}
which are indeed the Nambu equations of 
motion under the condition $\partial_3G=0$, 
as desired. 

The second resolution of the difficulty in step (III) 
is essentially to decompose the two-dimensional 
base space $(\xi^1,\xi^2)$ further into 
fiber and base space (1/1) 
$\bigl(\xi^1, \xi^2(\xi^1,t)\bigr)$ 
in which $\xi^2$ is now regarded as a new 
fiber direction and  $\xi^3$ is 
implicitly determined by setting the 
condition
\begin{align}
	\bar{H}(\xi^1,\xi^2, \xi^3(\xi^1,\xi^2;E)=E
\end{align}
where $E$ is a constant, as in the 
time-independent HJ formalism in the ordinary 
Hamiltonian dynamics. Thus the decomposition 
of the phase space is 
made in two steps $(3)\rightarrow (1/2)\rightarrow 
(1/1/1)$. 
This method can in principle be 
applied to arbitrary choices 
of two Hamiltonians $(H,G)$ without 
using the N-gauge symmetry. 
In this case again, the generalized HJ 
equations are reducible to an almost usual 
HJ formalism,  
but with more complicated relation between 
the momentum variable and the coordinate 
$\xi^2$ along the fiber direction.

As for the details of the other case of decompositions  
of the phase space into fiber and base space, 
namely, the (2/1) decomposition 
$\bigl(\xi^1, \xi^2(\xi^1,t), 
\xi^3(\xi^1,t)\bigr)$, the reader is 
recommended to consult the original paper \cite{yonegHJ}. 
Here we only mention that, in the language of 
differential forms, 
the generalized HJ system is characterized 
by the vanishing condition under projection 
for two independent 2-forms,  
instead of a single 3-form $\Omega^{(3)}$ of the 
(1/2) formalism,  
\begin{align}
	&\Omega^{(2)}_2\equiv 
	d\xi^2\wedge d\xi^1+ (\partial_3HdG-\partial_3GdH)
	\wedge dt, \\
	&\Omega^{(2)}_3\equiv 
	d\xi^3\wedge d\xi^1- (\partial_2HdG-\partial_2GdH)
	\wedge dt, 
\end{align}
which are related to the $\Omega^{(3)}$ by\footnote{
\eqref{3to2} has not been explicitly mentioned in \cite{yonegHJ}. We also warn the reader that there are 
trivial typos:  $d\xi^2\rightarrow d\xi^1$ in eq. (79), 
and $\partial_1\hat{H}=\partial_1\hat{G}=0
\rightarrow \partial_1\hat{H}=0$ in the first line in page 21.}
\begin{align}
	\Omega^{(3)}=-\Omega_2^{(2)}
	\wedge (d\xi^3-X^3dt)
	=\Omega_3^{(2)}
	\wedge (d\xi^2-X^2dt). 
	\label{3to2}
\end{align}
It seems that this pair of two 2-forms has never been 
mentioned in the literature. 
Geometrically,
 they are related to the 
Nambu equations of motion by the null conditions 
in the same sense as \eqref{null} for $\Omega^{(3)}$, 
\begin{align}
	i_L(\Omega^{(2)}_2)=i_L(\Omega^{(2)}_3) =0 .
	\label{nflow}
\end{align}
Using the language of the Einstein projection, 
these are equivalent with the vanishing 
conditions (=EE equations)
\begin{align}
	&\hat{\Omega}^{(2)}_2\equiv 
	-\partial_t\xi^2 d\xi^1\wedge dt+(\partial_3H\partial_1
	\hat{G}-\partial_3G\partial_1\hat{H})d\xi^1\wedge dt=0, 
	\\
	&\hat{\Omega}^{(2)}\equiv 
	-\partial_t\xi^3d\xi^1\wedge dt-
	(\partial_2H\partial_1\hat{G}-\partial_2G\partial_1
	\hat{H})d\xi^1\wedge dt=0, 
\end{align}
where $\hat{H}(\xi^1,t)\equiv 
H\bigl(\xi^1,\xi^2(\xi^2(\xi^1,t),\xi^3(\xi^1,t)\bigr)$ 
{etc.}
Actually, the final result of this case 
is essentially equivalent with the one 
we obtain through the second of the two methods in the 
(1/2) formulation above, as is expected since in that 
method the base coordinate is the same 
one-dimensional direction of $\xi^1$ in the end, 
corresponding to the final decomposition 
$(1/1/1)$. 

Of course, our discussion cannot exclude 
other possibilities for resolving the problem. 
 It might also be 
of some interest, quite independently of 
Nambu mechanics proper, 
to extend the above structure of 
the `vorticity bunlde' 
to new higher canonical structures 
 with 
multiple vorticity components for 
configuration spaces of arbitrary 
dimensions.  

\subsection{An example : the Euler top}
We exhibit an application of the formalism 
of the previous subsection, rather than pursuing 
the general theory further.\footnote{It is possible to 
extend the previous discussions to the Nambu equations 
of motion in general $n$-dimensional phase space, as discussed in the Appendix of 
\cite{yonegHJ}. } It would be 
useful for interested readers to get 
a concrete picture on what the generalized 
HJ theory for the Nambu mechanics is. 
Here again we restrict ourselves to the (1/2) formalism. 

In this case, we have 
$H=\frac{1}{2}\bigl((\xi^1)^2+(\xi^2)^2+(\xi^3)^2\bigr), 
G=\frac{1}{2}\bigl((\xi^1)^2/I_1+(\xi^2)^2/I_2+(\xi^3)^2/I_3\bigr)$. 
By performing an N-gauge 
transformation with the generator $\Lambda=H^2/(2I_3)$, 
we have a new set of Hamiltonians satisfying 
the gauge condition $\partial_3G=0$,
\begin{align}
	H\rightarrow H=\frac{1}{2}\bigl((\xi^1)^2+(\xi^2)^2+(\xi^3)^2\bigr)
	,
	\quad 
	G-H/I_3\rightarrow G=\frac{\alpha}{2}(\xi^1)^2
	+\frac{\beta}{2}(\xi^2)^2, 
\end{align}
where
\begin{align}
	\alpha\equiv \frac{I_3-I_1}{I_3I_1}, 
	\quad 
	\beta\equiv \frac{I_3-I_2}{I_3I_2}. 
\end{align} 
The generalized HJ equations are now reduced to 
\begin{align}
&\partial_tS=-\frac{1}{2}
\bigl((\xi^1)^2+(\xi^2)^2+(\xi^3)^2\bigr), \\
&
\xi^3=\partial_1G\partial_2S-\partial_2G\partial_1S=
\alpha \xi^1\partial_2S-\beta\xi^2\partial_1S. 
\end{align}
It is convenient to change the variables 
$(\xi^1,\xi^2)$ 
to the elliptic coordinate $(G,u)$,
using Jacobi's elliptic functions.\footnote{Our notations 
are such that ${\rm sn}^2\,u+{\rm cn}^2\,u=1, 
k\,{\rm sn}^2\,u +{\rm dn}^2\,u=1, 
{\rm sn}'\,u={\rm cn}\,u\,{\rm dn}\,u, 
{\rm cn}'\,u=-{\rm sn}\,u\, {\rm dn}\,u, {\rm dn}'\,u=
-k^2{\rm sn}\,u\, {\rm cn}\,u$, 
with abbreviation 
${\rm sn}\,u={\rm sn}(u;k)$ {\it etc} suppressing the 
modulus parameter $k$. 
}
\begin{align}
	\xi^1=\Bigl(\frac{2G}{\alpha}\Bigr)^{1/2}
	{\rm sn}\, u, 
	\quad 
	\xi^2=\Bigl(\frac{2G}{\beta}\Bigr)^{1/2}
	{\rm cn}\, u. 
\end{align}
The modulus parameter $k$ of the elliptic function 
will be fixed 
later such that the above HJ equation 
takes the simplest form. 
Note that, though originally $G$ is the radial coordinate, 
the HJ equation is a differential equation 
with respect only to the coordinate $u$ and 
hence $G$ can be treated as a constant parameter. Then 
\begin{align}
	\xi^3=-\frac{(\alpha\beta)^{1/2}}{{\rm dn}\,u}
	\partial_uS.
\end{align}
By defining the reduced time-independent 
function $\bar{S}$ as 
\begin{align}
	S=-E(t-u_0)+\bar{S}(\xi^1,\xi^2)
\end{align}
where $E$ and $t_0$ are constants, 
the HJ equation becomes
\begin{align}
	\frac{\alpha\beta}{2{\rm dn}^2\,u}(\partial_u\bar{S})^2
	=E+\frac{G(\alpha-\beta)}{\alpha\beta}
	{\rm sn}^2\,u-\frac{G}{\beta}.
\end{align}
Then we choose the modulus parameter k as 
\begin{align}
	k^2=\frac{G(\beta-\alpha)}{\alpha\beta E_0}, 
	\quad 
	E_0=E-\frac{G}{\beta}, 
\end{align}
and the HJ equation can be integrated,\footnote{A trivial typo of eq. (94) in \cite{yonegHJ} is corrected here. }
\begin{align}
	&\bar{S}=\frac{A}{\alpha\beta}{\cal E}(u), 
	\quad \xi^3=-\sqrt{2E_0}\,{\rm dn}\, u, 
	\quad A\equiv (2\alpha\beta E_0)^{1/2}, 
	\\
	&{\cal E}(u)\equiv \int_0^{{\rm sn}\, u}
	dx \Bigl(\frac{1-k^2x^2}{1-x^2}\Bigr)^{1/2},
\end{align}
where the last expression ${\cal E}(u)$ is 
known as the fundamental elliptic integral 
of second kind, obeying the differential equation, 
\begin{align}
	\frac{d{\cal E}}{du}={\rm dn}^2\,u 
	={\rm cn}\,u\, {\rm dn}\, u 
	\frac{{\rm dn}\,u}{{\rm cn}\,u}=
	\Bigl(\frac{1-k^2{\rm sn}^2\, u}{1-{\rm sn}\, u}
	\Bigr)^{1/2}\frac{d\,{\rm sn}\,u}{du}.
\end{align}
In terms of the original independent base-space 
coordinates $(\xi^1,\xi^2)$, the final result for a complete solution 
with two integration constants $E$ and $t_0$ is 
\begin{align}
	S=-E(t-t_0)+
	\frac{A}{\alpha\beta}
	\int_0^{\xi^1/\sqrt{(\xi^1)^2+
	\frac{\beta}{\alpha}(\xi^2)^2}}
	\Bigl(\frac{1-k^2x^2}{1-x^2}\Bigr)^{1/2}
	dx.
\end{align}

Let us check whether this solution yields the original 
Nambu equations of motion following the 
prescription derived in the previous subsection. 
The Jacobi condition is 
\begin{align}
	\frac{\partial S}{\partial E}={\rm constant}
\end{align}
together with the condition 
$\frac{dG}{dt}=0$. Since we have included 
$t_0$, the constant on the r.h.side 
can be absorbed into the definition of $\bar{S}$. 
Now taking the total time derivative, the 
Jacobi condition leads to 
\begin{align}
	\frac{d}{dt}\Bigl(
	\frac{\partial \bar{S}}{\partial E}\Bigr)=
	\frac{du}{dt}\partial_u
	\Bigl(\frac{\partial \bar{S}}{\partial E}\Bigr)=1.
\end{align}
By some straightforward calculation using 
various properties of the elliptic functions 
(see \cite{yonegHJ}), we find
\begin{align}
	u=At,
\end{align}
up to the an arbitrary choice of the origin of time $t$. 
This immediately gives the familiar 
general solution for the Euler top:
\begin{align}
	\xi^1=\sqrt{\frac{2G}{\alpha}}{\rm sn}\, At, 
	\quad 
	\xi^2=\sqrt{\frac{2G}{\beta}}{\rm cn}\, At, 
	\quad 
	\xi^3=-\sqrt{2(E-G/\beta)}
	{\rm dn}\, At.
\end{align} 
If we compare this derivation with the one  
using the ordinary HJ treatments, 
our method seems much more 
elegant even from a 
practical viewpoint putting aside the question 
of principle, since the components 
of angular momentum themselves are directly the canonical 
variables: the process of rewriting 
the system in terms of the Euler angles 
as canonical coordinates is 
completely circumvented. Perhaps for other systems that 
can be succinctly expressed in the 
framework of the Nambu mechanics, it seems natural to 
expect the same merit. 

\subsection{Implication to quantization}
The $(1/2)$ formulation outlined in subsection 5.3 
with the axial gauge condition $\partial_3G=0$ suggests 
the following quantum version, at least semi-classically. 
The Hilbert space consists of functions 
$\langle \xi^1,\xi^2|\psi_1(t)\rangle$ and 
$\xi^3$ is interpreted as a first-
order differential operator, `vorticity operator'
\begin{align}
	\xi^3\rightarrow -i\hbar (\partial_1G\partial_2
	-\partial_2G\partial_1)=-i\hbar d_G.
\end{align}
The Schr\"{o}dinger equation is 
\begin{align}
	i\hbar \langle\xi^1,\xi^2|\psi_1(t)\rangle
	=H(\xi^1,\xi^2, -i\hbar d_G)
	\langle \xi^1,\xi^2|\psi_1(t)\rangle, 
\end{align}
which automatically preserves $G$. 
 In the WKB approximation,  
$\langle \xi^1,\xi^2|\psi_1(t)\rangle 
\sim e^{iS(\xi^1,\xi^2,t)/\hbar}$ this 
reduces to the HJ equation \eqref{axialgHJ} 
with \eqref{xi3vortex}, 
as it should be. 

One might wonder whether and how this quantization 
is understood from the viewpoint of the 
canonical structure of Nambu bracket emphasized in 
section 2. That is easily seen if we recall, as 
has been already pointed out in \cite{takh}, 
the usual Poisson bracket structure is  
buried or subordinated 
in the Nambu bracket in the following 
sense. Define a 2-bracket
\begin{align}
	\{A,B\}_G\equiv \{A,G,B\}
\end{align}
 by treating 
one, say $G$, of the two Hamiltonians. 
The Jacobi identity for this 2-bracket is 
an automatic consequence of the FI identity. 
\begin{align}
	\{\{A,B\}_G,C\}_G+\{\{B,C\}_G,A\}_G+\{\{C,A\}_G,B\}_G=0.
\end{align}
If we assume the axial gauge condition $\partial_3G=0$, 
the canonical Nambu bracket is rewritten using 
this 2-bracket into 
\begin{align}
	\{\xi^1,\xi^2\}_G=0, 
	\quad \{\xi^3,\xi^1\}_G=-\partial_2G, 
	\quad \{\xi^3,\xi^2\}_G=\partial_1G,
\end{align}
and correspondingly the Nambu equations of motion to
\begin{align}
	\frac{d\xi^i}{dt}=\{H,\xi^i\}_G.
\end{align}
It is clear that by replacing the above 
2-bracket by a commutator $-i\hbar\{\,\, ,\,\}_G
\rightarrow [\,\, ,\,]$, we are naturally led to 
the above wave-mechanical quantization. 

Although we have not presented in this review the 
details of the alternative treatment of 
the case of 1-dimensional base space $\xi^1$ in 
terms of the $(1/1/1)$ or $(2/1)$ formalism, 
the final results in this case can also 
be explained on the basis of a subordinate 2-bracket 
in a similar way. The Hilbert space 
consists of functions $\langle \xi^1|\psi_2(t)\rangle$. 
We treat $H$ as a conserved quantity $H=E$ from the 
beginning and define 
$\{\,\,,\,\}_H\equiv \{H,A,B\}$ 
assuming $\partial_3H\ne 0$.\footnote{
If we consider the Euler top, this 2-bracket 
gives the usual algebra \eqref{angularPB}
of angular momenta. }
Then the canonical Nambu bracket takes the form
\begin{align}
	\{\xi^1,\xi^2\}_H=\partial_3H, 
	\quad \{\xi^3,\xi^1\}_H=\partial_2H, 
	\quad \{\xi^3,\xi^2\}_H=-\partial_1H, 
	\label{111-2bra}
\end{align}
and the Nambu equations of motion is
\begin{align}
	\frac{d\xi^i}{dt}=\{G,\xi^i\}_H. 
\end{align}
Under the constraint
\begin{align}
	H\bigl(\xi^1,\xi^2(\xi^1,t),\xi^3(\xi^1,t)\bigr)=E,
	\label{Econstraint} 
\end{align}
we also have
\begin{align}
	0=\partial_2H+\partial_3H\partial_2\xi^3, 
	\quad 
	0=\partial_1H+\partial_3H\partial_1\xi^3, 
\end{align}
which are equivalent with the last two 
2-bracket relation exhibited in \eqref{111-2bra}. 
The first one in \eqref{111-2bra} is, after 
quantization, interpreted as giving 
the expression of the operator $\hat{\xi^2}$ 
in terms of the differential operator 
$\partial_1$ 
\begin{align}
	-i\hbar \partial_1=-
	\int^{\hat{\xi^2}}\frac{dx}{\partial_3H(\xi^1,x)},
\end{align}
such that 
$
[\xi^1,\hat{\xi^2}]=-i\hbar \partial_3H(\xi^1,\hat{\xi^2})$.
The Schr\"{o}dinger equation is now 
\begin{align}
	i\hbar\langle\xi^1|\psi_2(t)\rangle=
	\bar{G}(\xi^1,\hat{\xi^2})
	\langle \xi^1|\psi_2(t)\rangle, 
\end{align}
where the symbol $\bar{G}$ means that 
$G$ is regarded as a function of 
the base-space coordinate and the 
differential operator $\hat{\xi^2}$ by 
eliminating $\xi^3$ implicitly through 
the constraint \eqref{Econstraint}.
The complexity of this operator is a 
price we have to pay, since we do not assume any 
special gauge condition 
with respect to the N-gauge symmetry. 

These two approaches to the quantization 
look quite different: even the dimensions 
of the base space are different from each other, 
the first 
being of two dimensions while the 
second of one dimension. 
And yet in the classical limit we must have 
the same Nambu equations of motion. 
Some of the readers may feel that our 
reliance on the 2-bracket structure 
is a backward step if one takes the 
viewpoint that two Hamiltonians 
$H,G$ should appear on an equal footing. 
However, the presence of the N-gauge 
symmetry of the Nambu equations of motion 
means that such a description with 
manifestly symmetrical appearance 
of both $H$ and $G$ necessarily 
has a large degree of arbitrariness 
which is usually foreign to a definite form of 
quantum theory, 
as we have learnt in the gauge theory or the 
quantum gravity: to have a definite 
formulation we always need to fix such 
gauge degrees of freedom to some extent. 
Otherwise we have to attain a description only 
in terms of completely gauge-invariant 
observables. That would be a dauntingly 
difficult task either, if one is not allowed 
to make any approximation, such as 
lattice approximation for the Yang-Mills theory as we have 
alluded to in the introductory part of this section. 

From these considerations, I emphasize that 
the real question of quantization in pursuing 
Nambu mechanics is how 
to make certain 
the invariance (or, more appropriately, {\it covariance}) 
of the dynamics and its 
physical interpretation under the 
N-gauge symmetry. It seems that, in most of 
past attempts,   
a sufficient 
attention has not been paid to this 
question. Our result for the generalized 
HJ equation \eqref{gHJ1} 
given in subsection 5.2 encompasses  
the N-gauge symmetry and exhibits the new
 S-gauge symmetry. 
Keeping the latter symmetry in the same sense 
as the N-gauge symmetry {\it without} introducing too much 
redundancy must be important from this viewpoint. 

\section{Concluding remarks}
My motivation for studying the Nambu mechanics 
has originally been a hope that it might 
be useful for finding clues in exploring possible 
new methods of expressing dynamics 
at the most fundamental microscopic level 
in string/M theory or quantum gravity. 
In this review, I have discussed 
three works which I have done 
along this line from a unified and, 
as far as possible, 
elementary (or pedagogical) viewpoint. 
I have focused on the streams  
of basic ideas: in section 3 and 4,
 I have explained on how 
an aspect of the Nambu mechanics 
can be useful for thinking about 
possible new symmetries higher than the 
usual gauge symmetries with which we are 
familiar in formulating fundamental 
interactions. Then in section 5,
 a possible new canonical structure behind 
 the Nambu mechanics is discussed 
from a different perspective, the Hamilton-Jacobi 
theory, which has been scarcely 
taken up previously. 
Although, at the present level 
of development, I could not provide any 
compelling reason for the relevance 
of our results from the viewpoint of my original 
motivation, I hope that 
the questions which I have been asking here 
would become relevant in the near future. 
An old truth may connect to a new truth, or 
 ``an  
old sake in a new cup" (a favorite saying of Nambu 
\cite{nambutalk}) may sometimes help us, like the 
case where Hamilton-Jacobi theory in the 
19th century connected to quantum mechanics in the 
20th century.
Perhaps, we should also learn more about Nambu's 
passion and imagination for physics!

\section*{Acknowledgements}

I am grateful to the organizers 
of the workshop ``Space-time topology behind formation of micro-macro 
magneto-vortical structure by Nambu mechanics", 
(Osaka City University, Sept. 28--Oct.1, 2020)  
for invitation and for 
giving me a nice opportunity of reconsidering 
the problems treated here. 
I would also like to thank Akio Sugamoto and 
Yutaka Matsuo for useful discussions 
in preparing my talk. 

\appendix
\section{Dirac's attempt in the 1950s: a vortical stream in terms of the gauge potential}
In this Appendex, I briefly review Dirac's old works  
related in some broad sense to the ideas discussed 
in the present text, especially to the vortical 
structure associated with the Nambu mechanics which is 
the main theme of this workshop. 
In the workshop, this addemdum was provided as a topic 
for one of brainstorming sessions, in the hope 
of casting a different light on our subject 
from a historical perspective. 

From the viewpoint of the general 
theory of volume preserving flows, 
the Nambu equations of motion correspond to the 
particular (Clebsch) form 
for the potential function in \eqref{gaugefield}, 
\begin{align}
	A_i-\partial_i\psi=H\partial_iG . 
\end{align}
Essentially the same form was studied in a quite 
different context by Dirac in the 1950s. 
Dirac was pursuing a new classical 
theory of electrons, aiming at, as a final goal,  
a formulation of 
quantum electrodynamics that 
is completely free from ultraviolet infinities. 
He was thinking that the trouble of the 
usual QED was not a fault in the general principles 
of quantization, but should be ascribed
 to a wrong classical theory on which the 
usual formulation of QED was based. 
In one of his many attempts at remedying the situation, 
he tried to give exact classical equations 
without requiring any assumptions about the 
structure of electrons. 
In \cite{Dirac1}, he proposed to describe 
electric charges with no dynamical 
variables explicitly corresponding to them. 
As a first example of such a possibility, he 
proposed to study the Maxwell theory in terms of only the 
electromagnetic field under the 
gauge condition, 
\begin{align}
	A_{\mu}A^{\mu}=-k^2 
	\label{gcondition}
\end{align} 
with the Lagrangian density 
\begin{align}
	L=-\frac{1}{4}F_{\mu\nu}F^{\mu\nu}+\frac{1}{2}\lambda
	(A_{\mu}A^{\mu}+k^2), 
\end{align}
where $k$ and $\lambda$ are undetermined constant. 
The Maxwell equations lead to the current 
\begin{align}
	j_{\mu}=-\lambda (\partial_{\mu}S+A_{\mu}),
	\label{current}
\end{align}
where $S$ appears by taking into account the gauge degrees of 
freedom $A_{\mu}\rightarrow A_{\mu}+\partial_{\mu}S$. 
Thus the above gauge condition is now
\begin{align}
	(\partial_{\mu}S+A_{\mu})(\partial^{\mu}S+A^{\mu})=-k^2, 
	\label{WKB}
\end{align}
which can be interpreted as the Hamilton-Jacobi 
equation for an electron of mass $m=ke$ in the field 
of the gauge potential $A_{\mu}$.
Interestingly, Nambu came across to the same result 
in 1968 \cite{nambugcondition}: he also discussed the QED with the 
same gauge condition \eqref{gcondition} and 
drew the same analogy with the HJ equation, 
without knowing 
Dirac's old work at the time of his writing. 
He mentioned \cite{Dirac1} only as a 
postscript to the paper at the end. 
Nambu's original intention was to utilize the above 
gauge condition in order to interpret 
photon as a Goldstone boson arising as 
a result of (superficial) spontaneous breakdown of Lorentz invariance, 
providing the existence of a non-vanishing expectation 
value $\langle j_{\mu}\rangle\ne 0$.
  
Now let us go back to Dirac's works in the 50s. 
He noticed 
a problem with his result: 
\eqref{current} implies that the velocity 
field of the current $j_{\mu}$ is equal to
\begin{align}
v_{\mu}=k^{-1}(\partial_{\mu}S+A_{\mu})
\end{align}
and thus the vector $kv_{\mu}-A_{\mu}$ is 
irrotational. Since in practice there are 
situations where this vector can be vortical, 
he generalized his argument in his next work  \cite{Dirac2} on this subject 
as follows. 

First he replaces the usual classical (Lorentz) equations 
of motion for an electron moving with velocity, 
$	k\frac{dv^{\mu}}{ds}=v_{\nu}F^{\mu\nu}, 
$
to a stream of electrons, by looking upon 
$v^{\mu}$ as a continuous field as 
functions of space-time coordinates:
\begin{align}
	kv^{\nu}\partial_{\nu}v^{\mu}=v_{\nu}F^{\mu\nu}. 
\end{align}
Since by definition $v_{\nu}v^{\nu}=1$, it follows that 
\begin{align}
	v_{\nu}\partial_{\mu}v^{\nu}=0  \quad 
	\rightarrow \quad v_{\nu}f^{\mu\nu}=0
	\label{a8}
\end{align}
where 
\begin{align}
	f_{\mu\nu}=F_{\mu\nu}+k(\partial_{\nu}v_{\mu}
	-\partial_{\mu}v_{\nu})
	=\partial_{\mu}(A_{\nu}-kv_{\nu})
	-\partial_{\nu}(A_{\mu}-kv_{\mu}).
\end{align}
\eqref{a8} allows us to conclude that there exits a vector $u^{\mu}$ 
such that 
\begin{align}
	\frac{1}{2}\epsilon^{\mu\nu\rho\sigma}f_{\mu\nu}
	=v^{\rho}u^{\sigma}-v^{\sigma}u^{\rho}. 
\end{align}

In this four-dimensional context, the identity 
\begin{align}
	\epsilon^{\mu\nu\rho\sigma}
\partial_{\rho}f_{\mu\nu}=0
\label{Bidentity}
\end{align}
corresponds to the condition 
of incompressibility $\partial_iD^i=0$, when the argument 
is adapted to the gauge potential $A_i$ of the 3-
dimensional flow. 
Using this condition, Dirac shows 
that the two vectors $u_{\mu}$ 
and $v_{\nu}$ lie in 
integrable two-dimensional surfaces, and 
hence the stream lines of the electricity obeying 
$dx^{\mu}/ds=v^{\mu}$ are 
contained in these surfaces. 
He next introduces two independent scalar 
functions $\xi$ and $\eta$ which are 
constant on these surface, 
\begin{align}
	v^{\sigma}\partial_{\sigma}\xi=
	u^{\sigma}\partial_{\sigma}\xi=
	v^{\sigma}\partial_{\sigma}\eta=
	u^{\sigma}\partial_{\sigma}\eta=0, 
\end{align}
and hence satisfy
\begin{align}
	\epsilon^{\mu\nu\rho\sigma}f_{\mu\nu}\partial_{\sigma}\xi=0,\quad 
	\epsilon^{\mu\nu\rho\sigma}f_{\mu\nu}\partial_{\sigma}\eta=0.
\end{align}
This ensures that 
\begin{align}
	f_{\mu\nu}=\alpha (\partial_{\mu}\xi\partial_{\nu}
	\eta-\partial_{\nu}\xi\partial_{\mu}\eta),
\end{align}
where $\alpha$ is some scalar function, 
satisfying 
\begin{align}
	\epsilon^{\mu\nu\rho\sigma}\partial_{\mu}\xi
	\partial_{\nu}\eta\partial_{\rho}\alpha=0, 
\end{align}
because of the identity \eqref{Bidentity}. 
This implies that three vectors 
$\partial_{\mu}\xi, \partial_{\nu}
\eta, \partial_{\rho}\alpha$ are coplanar, and  
thus $\alpha$ is a function of $\xi$ and $\eta$. 
Then we can redefine a new pair of 
scalar functions $h$ and $g$ such that the 
Jacobian is equal to $\alpha$, 
\begin{align}
	\frac{\partial(h,g)}{\partial(\xi,\eta)}=\alpha, 
\end{align}
which allows us to express 
\begin{align}
	f_{\mu\nu}=\partial_{\mu}h\partial_{\nu}g
	-\partial_{\nu}h\partial_{\mu}g
	=\partial_{\mu}(h\partial_{\nu}g)
	-\partial_{\nu}(h\partial_{\mu}g). 
\end{align}
This is equivalent to 
the promised representation with a suitable 
choice of gauge, 
\begin{align}
	A_{\mu}=kv_{\mu}+h\partial_{\mu}g, 
\end{align}
which can involve a vortical component 
as desired. 

The reader must be aware of interesting  
parallelisms between Dirac's arguments 
and Nambu's discussions for the gauge potential 
for volume-preserving flows. 
Dirac considered a stream of 
charges, while Nambu discussed a 
stream in 3-dimensional phase space, which is 
in general vortical. However, once we assume that 
there exist a field $v^j$ satisfying $F_{ij}v^j=0$ 
for the field strength $F_{ij}$, 
Dirac's argument from \eqref{a8} applies 
equally well to the 3-dimensional incompressible 
flows. That this condition is guaranteed can be 
convinced if we 
reverse his arguments. 
In the Dirac case, 
a stream line lies on two-dimensional planes in four 
dimensions 
characterized by two vectors $v^{\mu}$ and 
$u^{\mu}$. A stream line in the 
Nambu case lies in the 
intersections of two surfaces with 
constant $H$ and $G$.

We can also mention 
that due to the nature of vortical 
flows, the Hamilton-Jacobi 
theory suggested from their common analogy 
can not take the usual form 
and must somehow be generalized. There is thus  
some flavor of parallelism between Dirac's 
discussion and our attempt 
of a generalized HJ theory 
discussed in section 5. 

Actually Dirac himself abandoned this approach later, 
because of the difficulty of quantization, 
and proceeded to yet another
 new idea \cite{Dirac3} which may 
be regarded as a precursor to modern string theory, 
or perhaps more appropriately to the string picture 
that emerges in the strong coupling 
regime of lattice gauge theory. 
In this work, he tried to formulate QED in a 
manifestly gauge invariant fashion. 
But I must stop here, since that would bring us to a 
subject which is too far 
from this special section.


\begin{thebibliography}{99}
\bibitem{nambu1973} Y.~Nambu, Phys. Rev. D7, 2405 (1973), 10.1103/PhysRevD.7.2405
\bibitem{nambuising} Y.~Nambu, Prog. Theor. Phys. 
1, 1 (1950),10/1143/ptp/5.1.1.
\bibitem{fusimi} K.~Husimi and I.~Syozi, 
Prog. Theor. Phys. 5, 177 (1950), 10.1143/ptp/5.2.177.
\bibitem{nambutalk} Y.~Nambu, https://thmat8.ess.sci.osaka-u.ac.jp/Meeting2013/.
\bibitem{takh} L.~Takhtajan, Comm. Math. Phys. 
160, 295(1994),10.1007/BF02103278.
\bibitem{almy} H.~Awata, M.~Li, D.~Minic and T.~Yoneya, 
JHEP 02, 013(2001)[hep-th/9906248], 10.1088/1126-6708/2001/02/013.
\bibitem{y1} T.~Yoneya, Prog. Theor. Phys. 97, 949 (1997), 10.1143/ptp/97.6.949.
\bibitem{Schild} A.~Schild, Phys. Rev.D16, 1722 (1977), 
10.1103/PhysRevD.16.1722.
\bibitem{nambu2} Y.~Nambu, p. 1, 
{\it Quark Confinement and Field Theory}, eds. 
D. R. Stump and D. H. Weingarten, 1977, John Wiley 
\& Sons. 
\bibitem{Hoppe} J.~Hoppe, Soryushiron Kenkyuu 80-3, 145, 1989.
\bibitem{dhn} B.~de Wit, J.~Hoppe and H.~Nicolai, 
Nucl. Phys. B305, 545 (1988),10.1016/0550-3213(88)90116-2.
\bibitem{yo3} T.~Yoneya, p. 419, {\it Wandering in the fields}, eds. K.~Kawarabayashi and A.~Ukawa, 
1987, World Scientific.
\bibitem{yo2} T.~Yoneya, Prog. Theor. Phys. 103, 1081 
(2000), 10.1143/ptp/103.6.1081.
\bibitem{bfss} T.~Banks, W.~Fischler, S.~H.~Shenker and 
L.~Susskind, Phys. Rev. D55, 5112 (1997), 10.1103/PhysRevD.55.5112. 
\bibitem{yonecovmat} T.~Yoneya, JHEP 06, 058 (2016), 
10.1088/1126-6708/2016/06/058.
\bibitem{Suss} L.~Susskind, hep-th/9704080.\bibitem{sekiyone} Y.~Sekino and T.~Yoneya, 
Nucl. Phys. B570, 174 (2000), 10.1016/S0550-3213(99)00793-2.
\bibitem{hnsy} M.~Hanada, J.~Nishimura, Y.~Sekino and T.~Yoneya, Phys. Rev. Lett. 104, 151601 (2010), 
10.1103/physrevlett.104.151601.
\bibitem{sekiyone2} Y.~Sekino and T.~Yoneya, 
Nucl. Phys. B619, 22 (2001), 10.1016/S0550-3213(01)00546-6. 
\bibitem{sheikh} M.~M.~Sheikh-Jabbari, 
JHEP 0409:017,2004, 10.1088/1126-6708/2004/09/017.
\bibitem{yonegHJ} T.~Yoneya, Prog. Theor. Exp. Phys. 
2017, 023A01, 10.1093/ptep/ptx008. 
\bibitem{yonelattice} T.~Yoneya, Nucl. Phys. B183, 471 
(1981),10.1016/0550-3213(81)90145-0.
\bibitem{kastrup} H.~A.~Kastrup, Phys. Rep. 101, 1 (1983).
\bibitem{nambuHJstring} Y.~Nambu, Phys. Lett. 92B, 327 
(1980), 10.1016/0370-2693(80)90275-0.
\bibitem{ein1917} A.~Einstein, Köglich Pr. Akad. der Wissenschaften, Sitzungsberichte, 606 (1917).
\bibitem{ein2} A.~Einstein, Deut. Phys. Gesellshaft 19, 82 (1917). English translations in The Collected Papers of
Albert Einstein (Princeton University Press, Princeton, 1997), Vol. 6.
\bibitem{arnold} V.~I.~Arnold, {\it Mathematical 
Methods of Classical Mechanics}, Springer, 
N.Y., 1978. 
\bibitem{Dirac1} P.~A.~M.~Dirac, Proc. R. Soc. London 
A209,291 (1951).
\bibitem{nambugcondition} Y.~Nambu, 
Suppl. Prog. Theor. Phys. Extra Number, 190 (1968).
\bibitem{Dirac2} P.~A.~M.~Dirac, Proc. R. Soc. London 
A212, 330 (1952). 
\bibitem{Dirac3} P.~A.~M.~Dirac, Can. J. Phys. 33, 650 
(1955). 

\end{thebibliography}
\end{document}